\documentclass{aa}

\usepackage{natbib}
\usepackage{graphicx}
\usepackage{txfonts}
\usepackage{hyperref}
\hypersetup{
    colorlinks=true,   
    urlcolor=blue,     
    linkcolor=red,     
    citecolor=blue,
}
\usepackage{multirow}
\usepackage{booktabs}

\begin{document}
  \title{Comparison of multifrequency positions of extragalactic sources from global geodetic VLBI monitoring program and \textit{Gaia} EDR3}
  \titlerunning{Comparison of multifrequency positions of extragalactic sources}
  \author{N. Liu\inst{1,2}
     \and
          S. B. Lambert\inst{3}
          \and
          P. Charlot\inst{4}
          \and
          Z. Zhu\inst{1}
          \and
          J.-C. Liu\inst{1}
          \and
          N. Jiang\inst{1}
          \and
          X.-S. Wan\inst{1}
          \and
          C.-Y. Ding\inst{1}
          \thanks{Now at the Philips (China) Investment Co., Ltd.}
          }

  \institute{School of Astronomy and Space Science,
   		     Key Laboratory of Modern Astronomy and Astrophysics (Ministry of Education), Nanjing University, Nanjing 210023, P. R. China\\
              \email{[niu.liu;zhuzi;jcliu]@nju.edu.cn}
         \and
             School of Earth Sciences and Engineering, Nanjing University, Nanjing 210023, P. R. China
         \and
             SYRTE, Observatoire de Paris,
             Universit\'e PSL, CNRS, Sorbonne Universit\'e, LNE, Paris, France\\
             \email{sebastien.lambert@obspm.fr}
         \and
             Laboratoire d'astrophysique de Bordeaux, Univ. Bordeaux, CNRS, B18N, All\'ee Geoffroy Saint-Hilaire, 33615 Pessac, France\\
             \email{patrick.charlot@u-bordeaux.fr}
             }

  \date{Received; accepted}

  \abstract
   {
       Comparisons of optical positions derived from the \textit{Gaia} mission and radio positions measured by very long baseline interferometry (VLBI) probe the structure of active galactic nuclei (AGN) on the milli-arcsecond (mas) scale.
       So far, such comparisons have focused on using the $S/X$ band (2/8 GHz) radio positions but have not taken advantage of the VLBI positions that exist at higher radio frequencies, namely $K$ band (24 GHz) and $X/Ka$ band (8/32 GHz).
   }
   {
       We extend previous works by considering two additional radio frequencies ($K$ band and $X/Ka$ band) with the aim to study the frequency dependence of the source positions and its potential connection with the physical properties of the underlying AGN.
   }
   {
       We compared the absolute source positions measured at four different wavelengths,
       that is, the optical position from the {\it Gaia} Early Data Release 3 (EDR3) and the radio positions at the dual $S/X$, $X/Ka$ combinations and at $K$ band, as available from the third realization of the International Celestial Reference Frame (ICRF3), for 512 common sources.
       We first aligned the three ICRF3 individual catalogs onto the \textit{Gaia} EDR3 frame and compare the optical-to-radio offsets before and after the alignment.
       Then we studied the correlation of optical-to-radio offsets with the observing (radio) frequency, source morphology, magnitude, redshift, and source type.
   }
   {
   The deviation among optical-to-radio offsets determined in the different radio bands is less than 0.5~mas,
   but there is statistical evidence that the optical-to-radio offset is smaller at $K$ band compared to $S/X$ band for sources showing extended structures.
   The optical-to-radio offset was found to statistically correlate with the structure index.
   Large optical-to-radio offsets appear to favor faint sources but are well explained by positional uncertainty, which is also larger for these sources.
   We did not detect any statistically significant correlation between the optical-to-radio offset and the redshift.
  }
   {
    The radio source structure might also be a major cause for the radio-to-optical offset. 
    For the alignment of with the \textit{Gaia} celestial reference frame, the $S/X$ band frame remains the preferred choice at present.
   }

  \keywords{reference systems -- 
            astrometry --
            techniques: interferometric --
            quasars: general --
            catalogs}

\maketitle


\section{Introduction}     \label{sec:introduction}

   The frequency-dependent position of extragalactic objects is of interest in both astrometric and astrophysical fields, especially the position offset between the optical centroid and radio core.
    Results of studies of position frequency dependency can be used to improve accuracy of astrometric catalogues, for example, \citet{2010A&A...510A..10A,2011A&A...532A.115C,2013MNRAS.430.2797A,2014JGeod..88..575S}.
   The study of optical-to-radio offset also provides a probing of the structural properties of Active Galactic Nucleus (AGNs), such as the accretion disc and relativistic jet \citep[e.g.,][]{2013A&A...553A..13O,2019ApJ...871..143P}.

   Accurate positions at sub- milli-arcsecond (mas) are needed for studying the frequency-dependence of the source position, which was achieved exclusively by the very long baseline interferometry (VLBI).
   The arrival of \textit{Gaia} Early Data Release 3 \citep[\textit{Gaia} EDR3;][]{2016A&A...595A...1G,2021A&A...649A...1G} provides optical positions with a precision close to that of VLBI.
   The comparison between \textit{Gaia} and VLBI positions derived at dual-band $S/X$ band (2/8~GHz) shows the agreement (angular separation) on the level of 1~mas for most sources except for about 6\%--22\% of outliers, that is, significant \textit{Gaia}/VLBI offsets \citep{2016A&A...595A...5M,2018A&A...616A..14G,2017MNRAS.471.3775P,2017MNRAS.467L..71P,2017A&A...598L...1K,2017ApJ...835L..30M,2018AJ....155..229F,2019MNRAS.482.3023P,2019ApJ...871..143P,2019ApJ...873..132M,2020MNRAS.493L..54K,2020A&A...644A.159C}.
   Recently, \citet{2019MNRAS.482.3023P} reported that for 62\% of sources with a significant \textit{Gaia}/VLBI offset and also a determinable jet direction, the VLBI-to-\textit{Gaia} offset vector is parallel to the jet.
   \citet{2019ApJ...871..143P,2020MNRAS.493L..54K} further studied those offsets and found correlations between the \textit{Gaia}/VLBI offset parallel to jet direction and the dominance of different AGN components in the optical emission, AGN types, and optical polarization properties.

   These studies, however, are only limited to the VLBI positions at dual $S/X$ band.
   We note that VLBI positions at higher frequencies, namely at $K$ (24~GHz) and dual $X/Ka$ band (8/32~GHz), are also of interest since they have precisions on the same order as those at $S/X$ band \citep[][]{2020A&A...644A.159C}.
   \citet{2002ivsg.conf..350J} suggested that the $K$ and $X/Ka$ band observations may suffer less from the radio source structure effects than those at the $S/X$ band, while \citet{2010AJ....139.1713C} found that the sources are more compact at the higher frequencies.
   Including the $K$ and $X/Ka$ band positions in the \textit{Gaia}/VLBI offset studies would help understand the origin of the optical-to-radio offsets.
   On the other hand, the alignment between $K$ and $X/Ka$ band VLBI catalogs and the \textit{Gaia} celestial reference frame (\textit{Gaia}-CRF) also requires detailed studies on the position offsets among the $K$ band, $X/Ka$ band, and \textit{Gaia} catalogs.

   We aim to compare the multifrequency positions of extragalactic sources to complement findings by \citet{2019MNRAS.482.3023P}.
   For this purpose, we computed the \textit{Gaia}-to-VLBI offsets at the $S/X$, $K$, and $X/Ka$ band and studied their dependency on the properties of extragalactic sources, such as the magnitude, redshift, and morphological properties.
   These comparisons are intended to provide new insights into the understanding of the origin of optical-to-radio offsets, and should help with the alignment of the \textit{Gaia}-CRF and VLBI frames other than at $S/X$ band.


\section{Materials and methods}    \label{sec:obs}

    \begin{table}[htbp]
        \centering
        \caption{\label{tab:median-err}
            Median formal uncertainties for 512 common sources among ICRF3 ($S/X$, $K$, and $X/Ka$ band) and \textit{Gaia} EDR3 catalogs.
        }
        \begin{tabular}{cccc}
        
            \hline \noalign{\smallskip}
            Catalog &$\sigma_\alpha\cos\delta$  &$\sigma_\delta$  &$\sigma_{\rm pos,max}$\\
            & $\mathrm{\mu as}$ & $\mathrm{\mu as}$  & $\mathrm{\mu as}$ \\
            \noalign{\smallskip}
            \hline
            \noalign{\smallskip}
            ICRF3 $S/X$          & 45  & 57  & 58  \\
            ICRF3 $K$            & 68  &132  &134  \\
            ICRF3 $X/Ka$         & 68  & 99  &107  \\
            \textit{Gaia} EDR3   &146  &122  &161  \\
            \hline
        \end{tabular}
        \tablefoot{$\sigma_{\rm pos,max}$ represents the semi-major axis of the error ellipse.
        }
    \end{table}

   We used the radio positions of sources at $S/X$, $X/Ka$ combinations and at $K$ band from the ICRF3 catalog \citep{2020A&A...644A.159C}, based on analysis of VLBI observations.
   For positions of their optical counterparts, we took the AGN sample (\texttt{gaiaedr3.agn\_cross\_id} table) in the \textit{Gaia} EDR3 from the \textit{Gaia} archive\footnote{\url{http://gea.esac.esa.int/archive/}},
   from where we found optical counterparts for 3181 ICRF3 sources via the external catalog name (column ``\texttt{catalogue\_name}'') for identifying these sources.
   The cross-match among these four catalogs gave a sample of 512 common sources.
    
   The median formal uncertainties in right ascension, declination, and along the semi-major axis of error ellipse in four catalogs are given in Table~\ref{tab:median-err}.
   For the sample used here, the median uncertainty of the $S/X$ band position is generally twice smaller than the $K$ and $X/Ka$ band ones, a property already noted by \citet{2020A&A...644A.159C} when intercomparing the sources common to the three ICRF3 catalogs. 
   This differs from when considering all sources in each catalog because the $S/X$ band catalog has a majority of survey sources which have lower position uncertainty. 
   Likewise, the $S/X$ band uncertainties are nearly four times better than the \textit{Gaia} EDR3 ones.

   We assumed that all catalogs may have distortions, and first wanted to remove them.
   We used the vector spherical harmonics \citep[VSH;][]{2012A&A...547A..59M} of degree 2 and followed similar procedures as those described in \citet{2020A&A...634A..28L}. 
   The VSH technique decomposes a vector field on the sphere onto a set of orthogonal vector functions in order to evince the features of the vector field at different scales.
   The first two degrees of VSHs model the large-scale differences between catalogs such as the orientation offset and declination-dependent systematics, thus they are useful for the purpose here.
   The \textit{Gaia} EDR3 position was chosen as the reference.
   By doing so, we can analyze multiwavelength positions in a framework as consistent as possible and avoid as much as possible bias arising from the alignment errors and deformations of the celestial reference frames.

   We then calculated three optical-to-radio offset quantities, namely, the angular separation $\rho$ between the $S/X$, $K$, or $X/Ka$ band positions and the \textit{Gaia} positions.
   We also computed the statistics of normalized separation, noted as $X$, following the same procedures as those in the \citet[][]{2016A&A...595A...5M},
   to account for the uncertainty and correlation between right ascension and declination of individual sources.
   These two quantities serve as indicators of significant optical-to-radio distance as done in recent works \citep[e.g.,][]{2018A&A...616A..14G,2019MNRAS.482.3023P}.
   
   In order to understand the origin of the optical-to-radio offsets, we studied their dependency on source properties, including the source morphology, magnitude, as well as redshift.
   We used the Pearson test without any a priori assumption to quantify the significance of the correlation.
   
   The source morphological property in the radio domain can be characterized by the structure index \citep[SI;][]{1997ApJS..111...95F,2015AJ....150...58F} at $S/X$ band, as available from the Bordeaux VLBI Image database (BVID)\footnote{\url{http://bvid.astrophy.u-bordeaux.fr/}}.
   The structure index is derived from the median value of the additional group delay for all Earth-based VLBI baselines due to the non-pointlike structure.
   It indicates the compactness of source: the larger the structure index value, the more extended the source.
   We used the median value of SI for each source.
   Since the BVID does not provide the values of SI for all sources in our sample, we also retrieved $X$ band images from the Astrogeo VLBI FITS image database\footnote{\url{http://astrogeo.org/vlbi_images/}} for the sources not in BVID and calculated the structure index following the same pipeline as that in BVID for those sources.
   
   We used the $G$ magnitude \citep[wavelength range 330--1050~nm;][]{2016A&A...595A...1G} given in the {\it Gaia} EDR3 catalogs.
   As the \textit{Gaia} position uncertainty degrades for fainter objects, the correlation between optical-to-radio offsets and the $G$ magnitude will indicate how the \textit{Gaia} uncertainty influences the optical-to-radio offsets.
   It is also aimed to examining whether optically-bright objects (e.g., with magnitude < 18) would be preferable to align the ICRF3 and \textit{Gaia}-CRF, as investigated in \citet{2008A&A...490..403B}.

   We were also curious about a possible dependence of the offset with redshift, as suggested in \citet{2014AJ....147...95Z,2017ApJ...835L..30M}.
   In order to check this effect, we included in our analyses the redshift $z$ taken from the the fifth release of the Large Quasar Astrometric Catalogue \citep[LQAC-5;][]{2019A&A...624A.145S}.

\section{Impact of systematics on optical-to-radio vector}    \label{sec:systematics-r2o}

    %
    \begin{figure*}[hbtp]
        \centering
        \includegraphics[width=60mm]{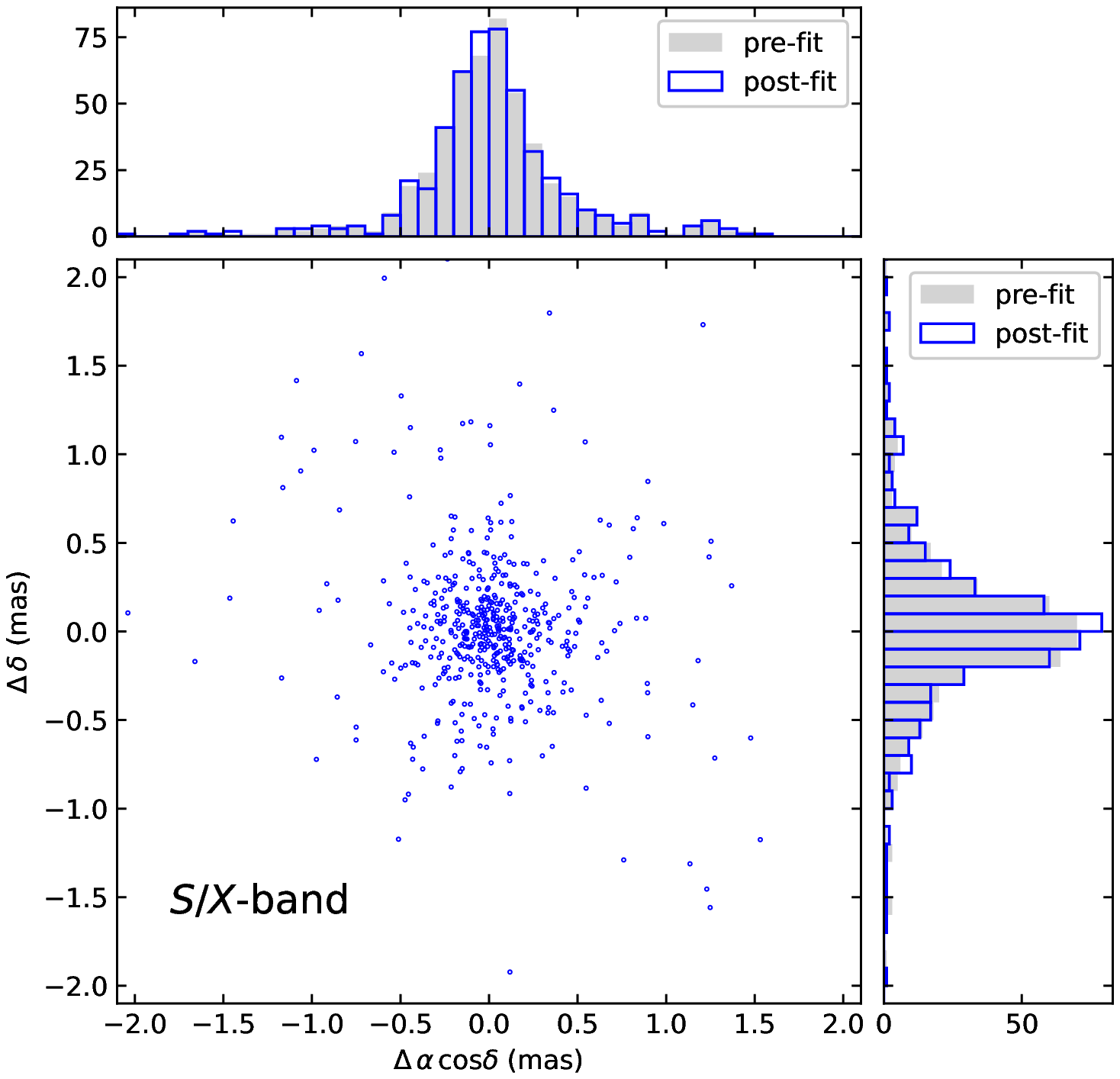}
        \includegraphics[width=60mm]{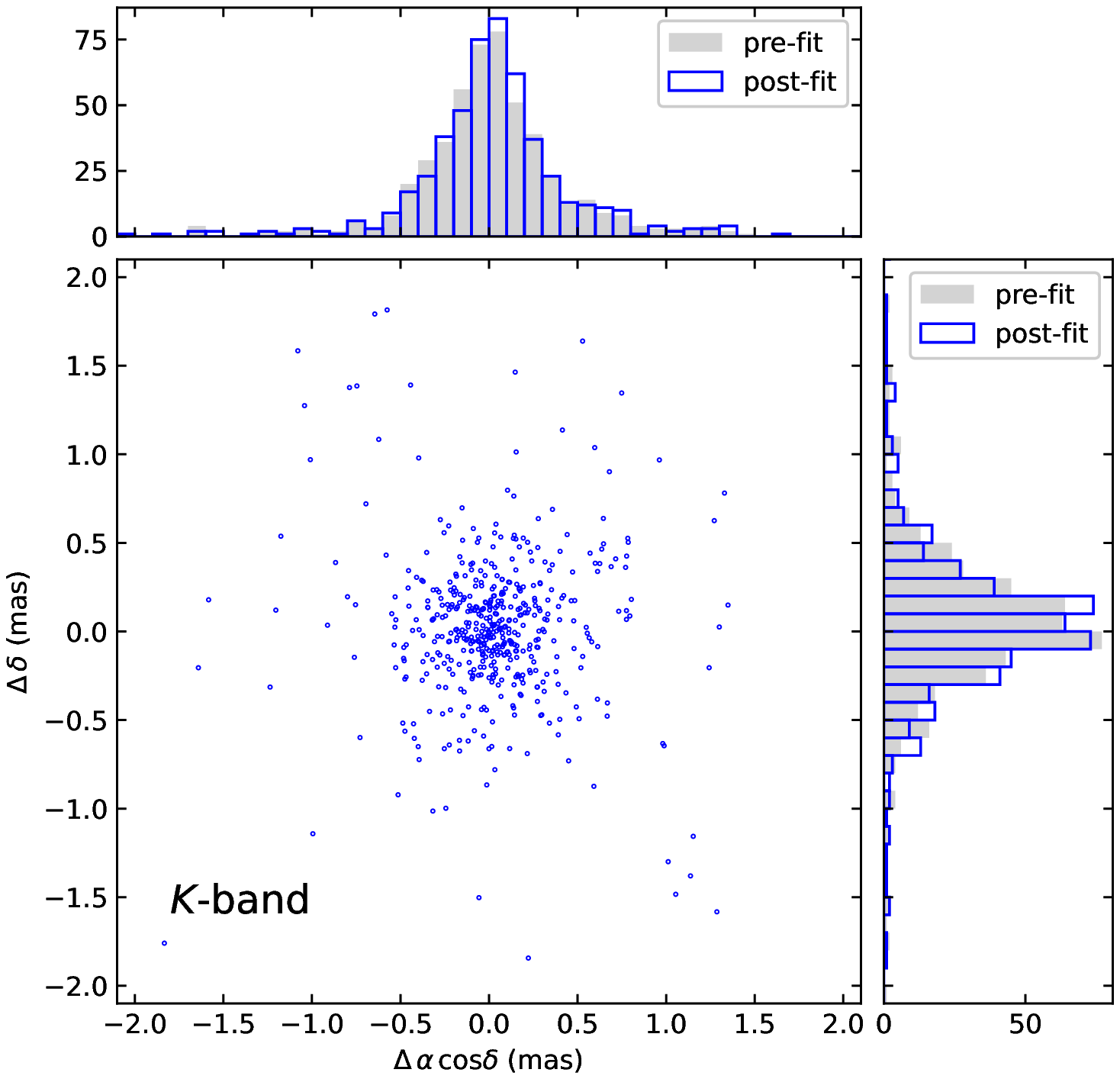}
        \includegraphics[width=60mm]{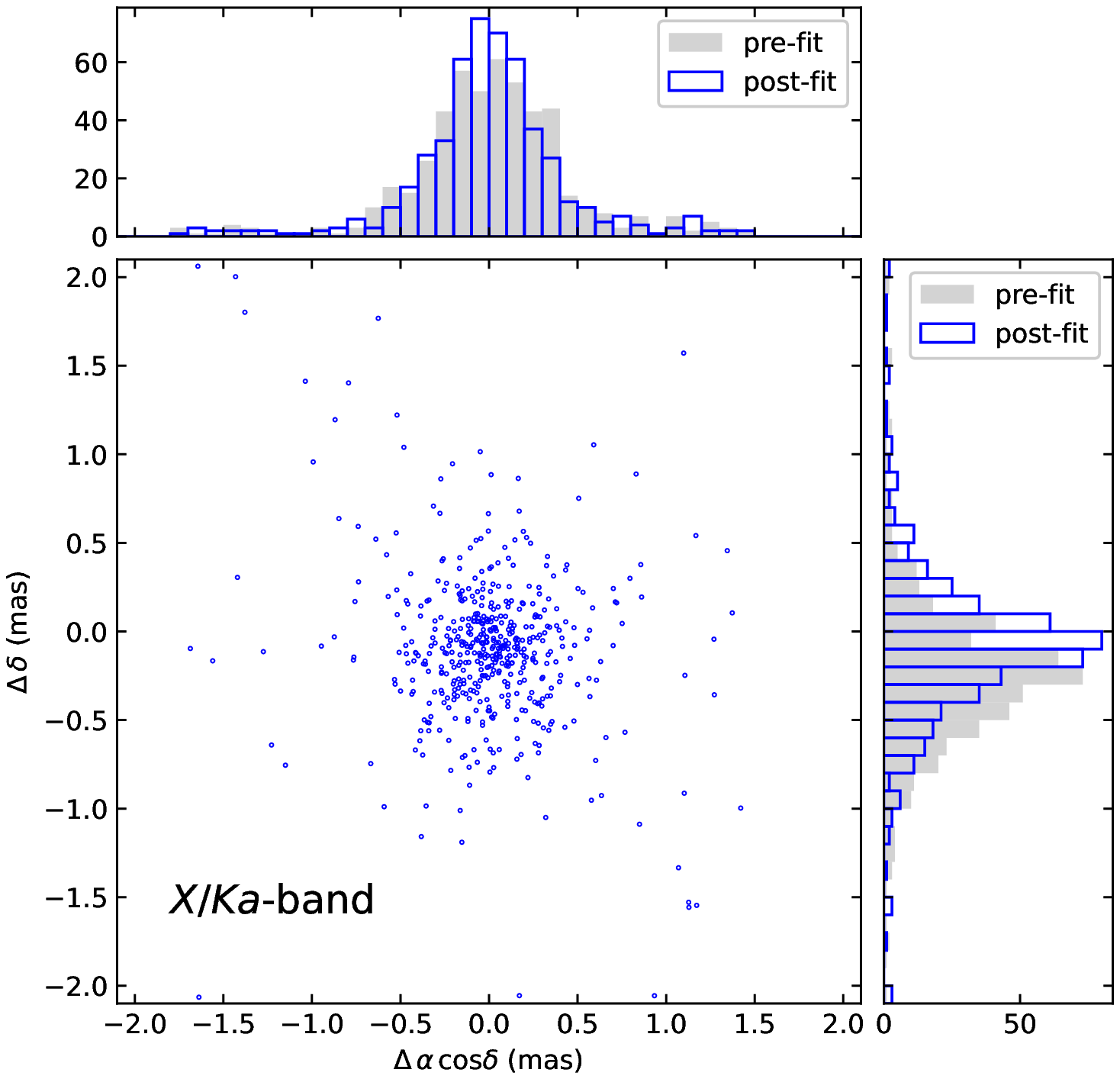}
        \caption[]{\label{fig:pos-offset-scatter}
            Scatter of the ICRF3 $S/X$ band (\textit{left}), $K$ band (\textit{middle}), and $X/Ka$ band (\textit{right}) positions with respect to the \textit{Gaia} EDR3 positions for the 512 sources in common, after removing relative deformations between these frames.
            The histograms in the sub-panels at the top and right side present the distribution of optical-to-radio differences in right ascension and declination before (pre-fit in grey) and after (post-fit in blue) applying the VSH transformation.
        }
    \end{figure*}
   
    \begin{figure*}[hbtp]
        \centering
        \includegraphics[width=50mm]{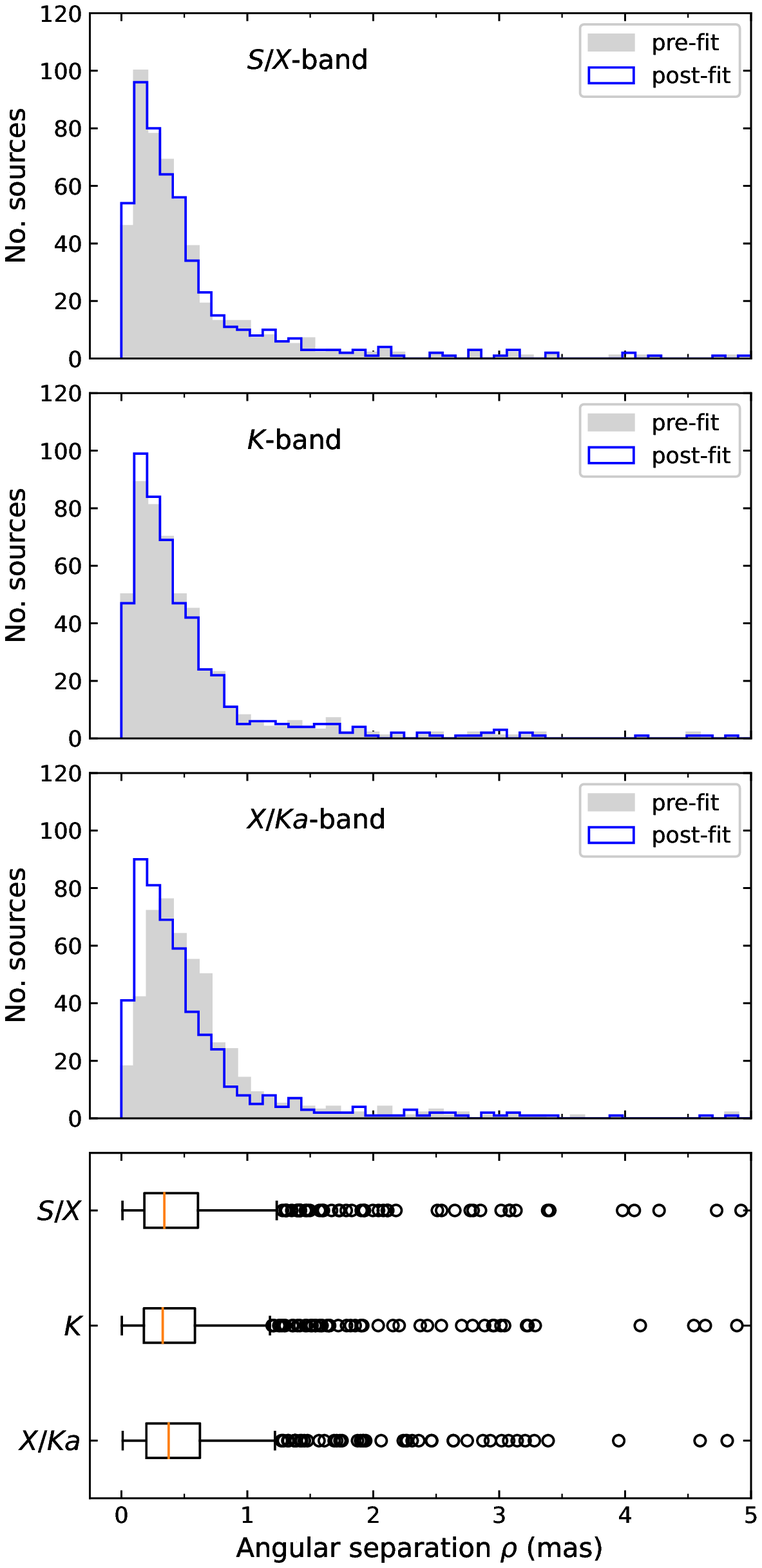}
        \includegraphics[width=50mm]{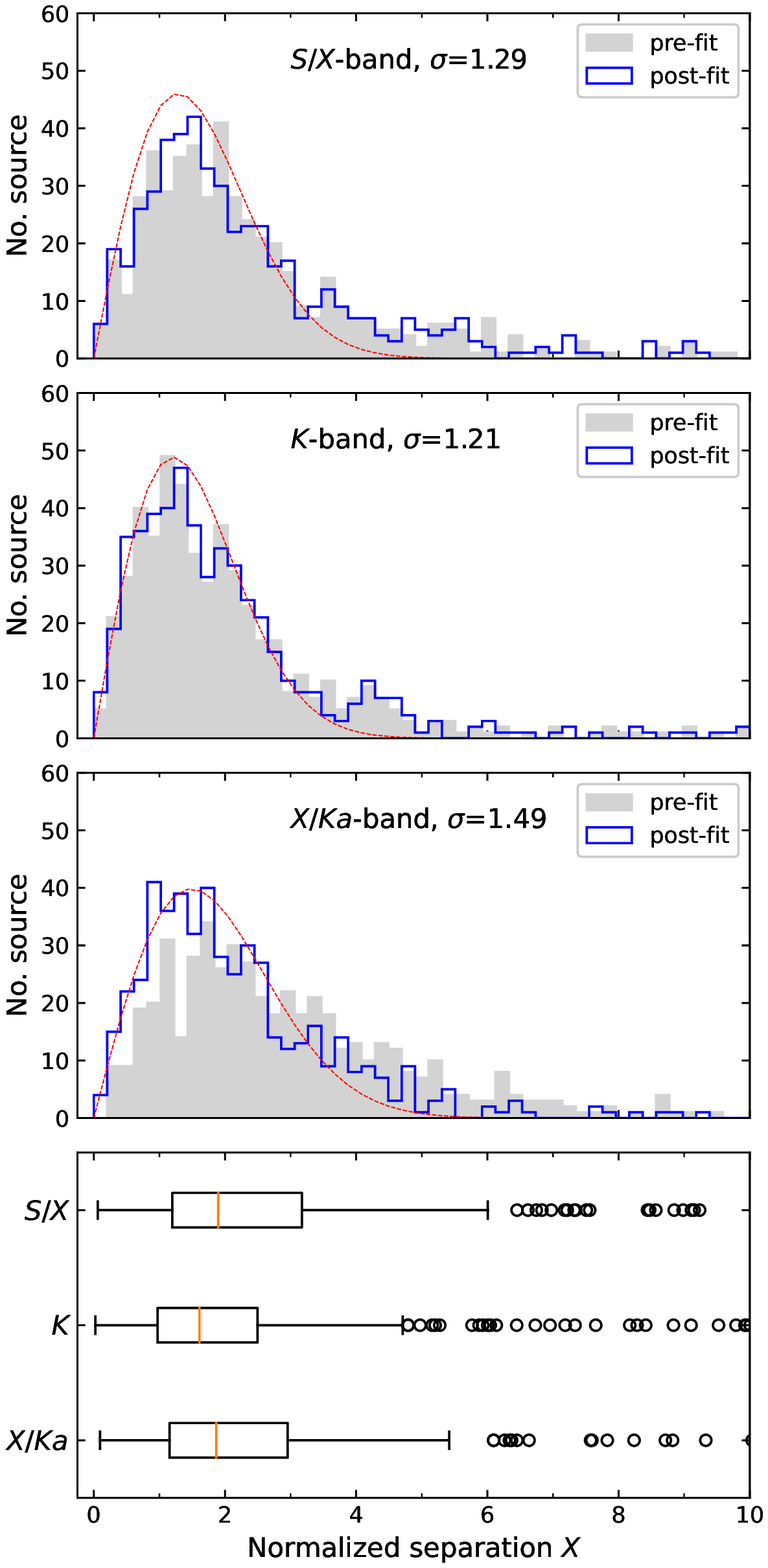}
        \caption[]{\label{fig:rho-hist}
            Distributions of optical-to-radio offset $\rho$ (\textit{left}) and normalized separation $X$ (\textit{right}) at $S/X$, $K$, and $X/Ka$ band for the 512 sources in common.
            The distribution is given for two cases, before (pre-fit in grey) and after (post-fit in blue) applying the VSH transformation.
            The left and right ends of box in the boxplot indicate the 25th and 75th percentiles (labelled as $Q_1$ and $Q_3$), respectively, while the orange line within the box shows the location of median value.
            The ends of the whisker on the left and right side show the lower and upper limits of the bulk of the sample, which are $Q_0=Q_1-1.5\times{\rm IQR}$ and $Q_4=Q_3+1.5\times{\rm IQR}$, where ${\rm IQR}=Q_3-Q_1$.
            Data points (open circle) greater than $Q_4$ or smaller than $Q_0$ are outliers suggested by the boxplot, of which the number is 50, 53, and 49 for $S/X$, $K$, and $X/Ka$ band, respectively, based on the angular separation; they are 26, 35, and 42 based on normalized separation.
            The red dashed lines in the first three plots on the right represent the Rayleigh distributions with the sigma to be best fitted to the sample of sources with $X<X_0$, where $X_0=3.53$ (Sect.~\ref{subsec:r2o-freq}). 
            There are 23 sources at $S/X$ band, 8 at $K$ band, and 12 at $X/Ka$ band with $X>10$, which are beyond the axis.
        }
    \end{figure*}

    Most of the VSH transformation parameters between the ICRF3 and $\textit{Gaia}$-CRF3 catalogs were in a range of 10--50\,$\mathrm{\mu as}$, except $D_3$ and $M_{\rm 20}$ for $X/Ka$ versus \textit{Gaia} that are $-205~\pm~30$ and $+153~\pm~35~\mathrm{\mu as}$.
    (The $D_3$ term represents a dipolar deformation along the Z-axis, while the $M_{\rm 20}$ term causes a deformation of $\sin 2\delta$ to the declination, respectively.)
    These three terms most likely reflect the zonal errors in the $X/Ka$ catalog due to weak geometry and sparse observations, as noted by \citet{2020A&A...644A.159C}. 
    For more discussions on these systematics, see \citet{2020A&A...634A..28L,2020A&A...644A.159C}.
    Here we were concerned about how these systematics affect the optical-to-radio offset.
    We compared distributions of optical-to-radio offset before and after applying the VSH transformation, denoted as ``pre-fit'' and ``post-fit'' cases, respectively.
    The optical-to-radio offset studied in Sect.~\ref{subsec:r2o-freq}-\ref{sec:r2o-corr}, if not specified, is referred to the post-fit one.
 
    Figure~\ref{fig:pos-offset-scatter} presents the post-fit offset scatter of $S/X$ band, $K$ band, and $X/Ka$ band positions relative to the \textit{Gaia} positions, including the distribution of scatter in right ascension and declination, for the 512 common sources.
    The agreement between ICRF3 and \textit{Gaia} positions is at the level of 0.3--0.5~mas for both coordinates.
    The distributions of optical-to-radio offsets in right ascension and declination are given for both the pre-fit and post-fit cases.
    Only marginal differences are found between the two cases for the ``$S/X\,-\,$\textit{Gaia}'' and ``$K\,-\,$\textit{Gaia}'' comparison.
    However, there is a declination bias of about 0.3~mas for the ``$X/Ka\,-\,$\textit{Gaia}'' comparison, in line with the systematics of the $X/Ka$ band catalog noted above, which seems to be corrected by the VSH transformation.
    
    The distributions of optical-to-radio offsets and normalized separations at $S/X$, $K$, and $X/Ka$ band are shown in Fig.~\ref{fig:rho-hist}.
    The pre-fit distribution of optical-to-radio offsets does not differ much from the post-fit one for $S/X$ and $K$ band but is generally shifted rightwards by about 0.1--0.2~mas for $X/Ka$ band.
    Similar to the case of optical-to-radio offsets, the pre-fit distribution of normalized separations is similar to the post-fit one for $S/X$ and $K$ band but its tail end (i.e., towards larger normalized separations) is thicker for $X/Ka$ band.

\section{Dependency of optical-to-radio offsets on observing frequency}    \label{subsec:r2o-freq}

    \begin{figure*}[hbtp]
        \centering
        \includegraphics[width=60mm]{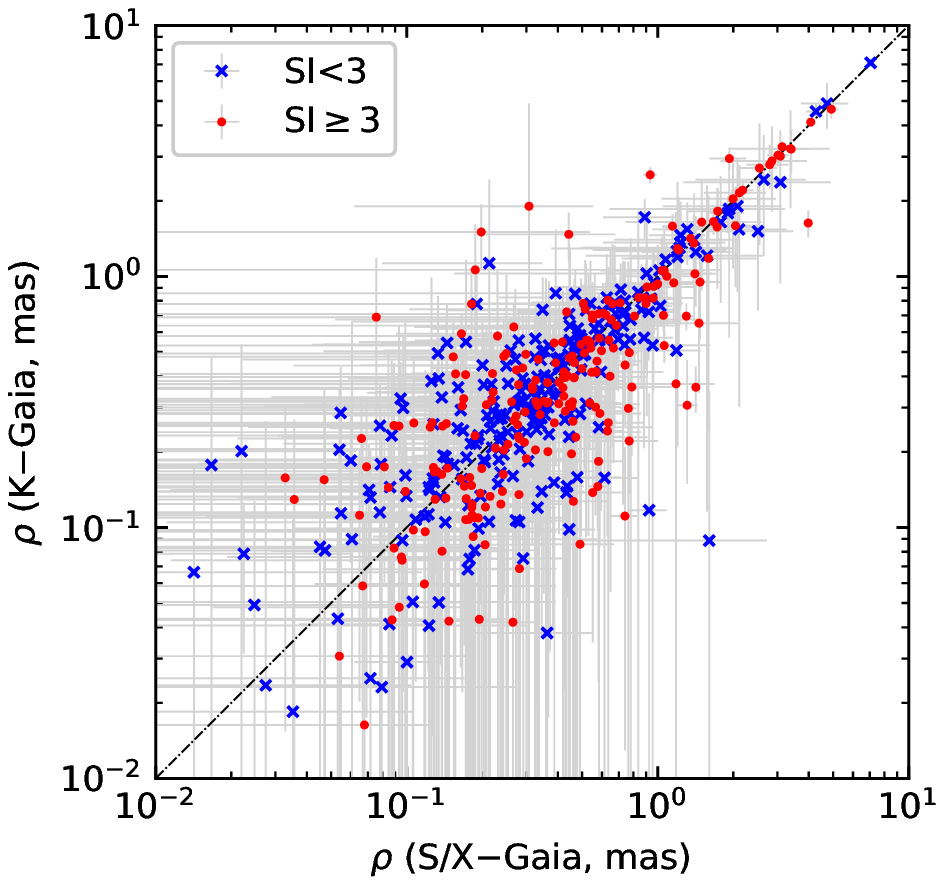}
        \includegraphics[width=60mm]{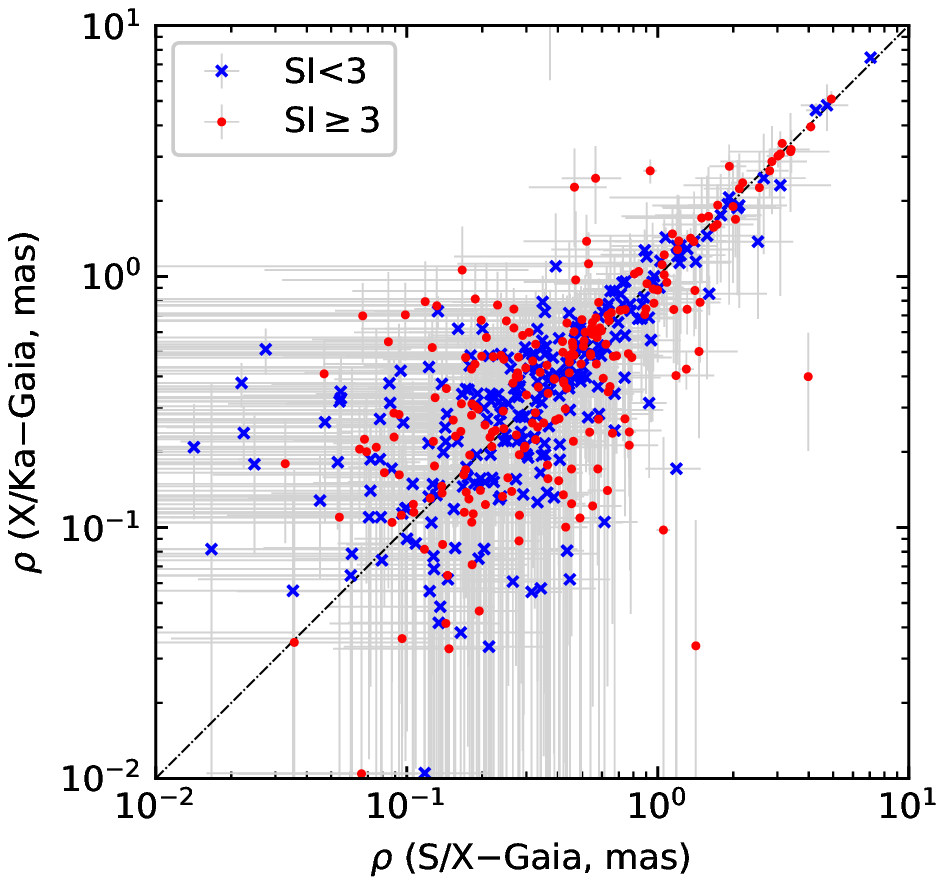}
        \includegraphics[width=60mm]{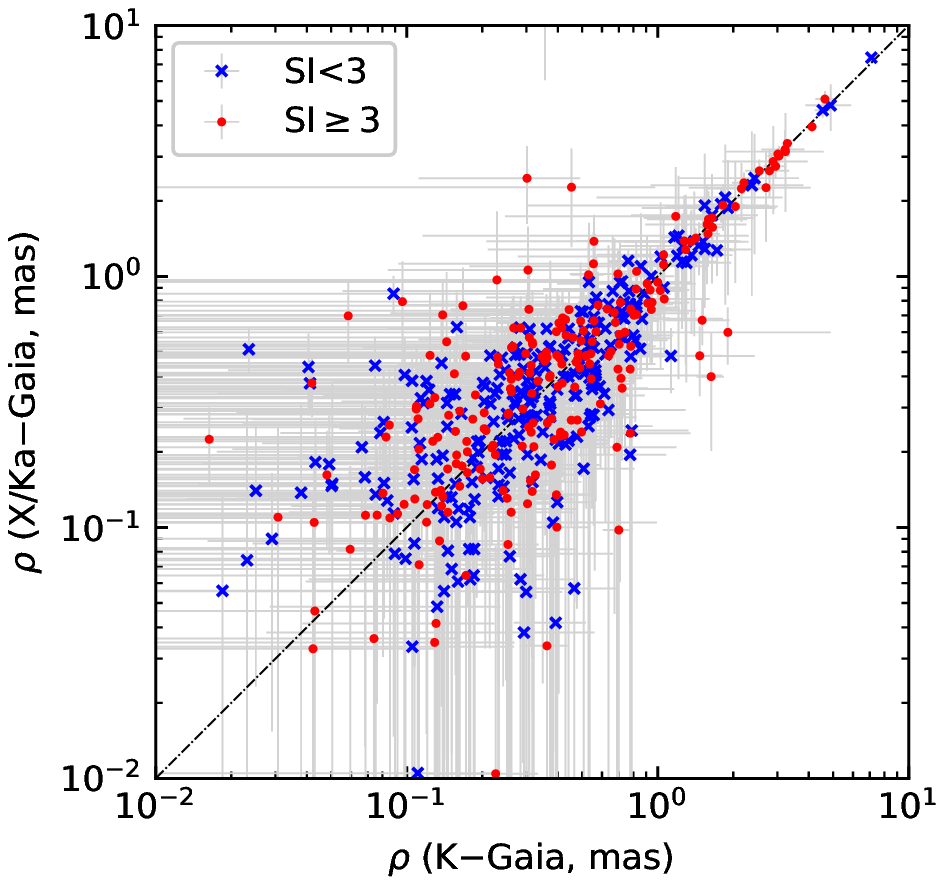}
        \caption[]{\label{fig:rho-com}
        Comparison of optical-to-radio distance between at the higher and lower frequencies for 512 sources common to the ICRF3 and \textit{Gaia} EDR3 catalogs. 
        Left: $K$ band vs. $S/X$ band;
        Middle: $X/Ka$ band vs. $S/X$ band;
        Right: $X/Ka$ band vs. $K$ band.
        The blue crosses and red circles distinguish sources with SI values less than 3 and greater than 3, respectively (SI $\geq 3$ suggests an extended structure).
        }
    \end{figure*}
    
    We compared the distribution of optical-to-radio offsets at different radio bands.
    As shown in Fig.~\ref{fig:rho-hist}, the distributions of the optical-to-radio offset $\rho$ at $S/X$, $K$, and $X/Ka$ band yield similar shapes, peaking at 0.1--0.2~mas.
    The number of outliers given by the boxplot, that is, optical-to-radio offsets greater than $\sim$1.2~mas\footnote{In the boxplot, the interquartile range (IQR) is defined as the distance between the 25th percentile ($Q_1$) and 75th percentile ($Q_3$), i.e., ${\rm IQR}=Q_3-Q_1$. Data points smaller than $Q_0=Q_1-1.5\times{\rm IQR}$ or greater than $Q_4=Q_3+1.5\times{\rm IQR}$ are considered as outliers. Since there is no outliers (open cirles) in the left side of boxplots in Fig.~\ref{fig:rho-hist}, we considered exclusively the upper limit $Q_4$, which is about 1.2~mas.} , is 50, 53, and 49 for $S/X$, $K$, and $X/Ka$ band, respectively.
    After removing the outliers, the 25th, 50th, and 75th percentiles are 0.20~mas, 0.38~mas, and 0.63~mas for the $S/X$ band; 0.18~mas, 0.33~mas, and 0.58~mas for the $K$ band; and 0.18~mas, 0.34~mas, and 0.61~mas for the $X/Ka$ band.
    
    In the right panel in Fig.~\ref{fig:rho-hist}, we plotted the distributions of normalized separation $X$  to check the significance of the optical-to-radio offsets.
    The distributions of $X$ are slightly different from those of $\rho$: it is sharper for the $K$ band but flatter for the $S/X$ band.
    There are 23 sources at $S/X$ band, 8 at $K$ band, and 12 at $X/Ka$ band with $X>10$, which are beyond the axis.
    For ideal cases, $X$ is supposed to follow a Rayleigh distribution of unit standard deviation.
    Medians of normalized separations are 1.89, 1.61, and 1.87 for $S/X$, $K$, and $X/Ka$ band, respectively, all greater than the predicted median value 1.18 for a standard Rayleigh distribution.

    We fitted the normalized separation distributions to the Rayleigh curve with an unknown of standard deviation $\sigma$.
    The fitting returned $\sigma$ values of 3.29 for $S/X$ band, 2.42 for $K$ band, and 2.62 for $X/Ka$ band, indicating that the \textit{Gaia}-to-$S/X$ band offset is more significant than the \textit{Gaia}-to-$K$ band and \textit{Gaia}-to-$X/Ka$ band offsets.
    Considering the bias introduced by the outliers in the fitting, we removed sources with large $X$ values and redid the fitting.
    Outliers were identified based on the prediction for a standard Rayleigh distribution.
    For a sample of $N$ sources with $X$ following a standard Rayleigh distribution, the number of sources with $X>X_0$ is expected to be less than one when $X_0=\sqrt{2\log{N}}$.
    For our sample ($N=512$), $X_0$ is 3.53.
    This criterion ruled out 115 sources for the $S/X$ band, 73 for the $K$ band, and 93 for the $X/Ka$ band.
    The percentage of outliers for the $S/X$ band corresponds to 22\% of the sample. 
    Interestingly, this percentage is the same as that found by \citet{2020A&A...644A.159C} when comparing the ICRF3 $S/X$ band frame and the \textit{Gaia}-CRF2 frame.
    The standard deviations for the ``clean'' sample then became 1.29, 1.21, and 1.49 for the $S/X$, $K$, and $X/Ka$ band, respectively.

    We wanted to test whether the optical-to-radio distance decreases at high frequency.
    For this purpose, we plotted the optical-to-radio offset of individual sources calculated at $K$ band and $X/Ka$ band against that calculated at $S/X$ band, and the optical-to-radio offset calculated at $X/Ka$ band against that calculated at $K$ band (Fig.~\ref{fig:rho-com}).
    Except for a small fraction (7\%), differences between the optical-to-radio offsets derived from the $S/X$ and $K$ band positions are generally smaller than 0.5~mas, and the same applies when comparing $X/Ka$ band to $S/X$ band, and the $X/Ka$ band to the $K$ band.
    We further performed sign tests between $S/X$ and $K$ band, $S/X$ and $X/Ka$ band, and $K$ and $X/Ka$ band.
    The null hypothesis is that the optical-to-radio offset is smaller at high frequency than at low frequency.
    To check this hypothesis, we counted the number of sources for which the optical-to-radio offset is smaller at $K$ band than at $S/X$ band, and we did the same for $X/Ka$ band vs. $S/X$ band and for $X/Ka$ band vs. $K$ band. 
    The numbers found in the three cases are 263, 239, and 237, respectively.
    Assuming that the count follows a binomial random distribution, the corresponding confidence levels to accept the null hypothesis are 75\%, 7\%, and 5\%, respectively.
    Considering that not all sources in our sample show significant extended structure, we performed the same sign test on the 228 sources that have an $X$ band structure index greater than 3 (Sect.~\ref{subsec:r2o-si}).
    The number of sources for which the high-frequency optical-to-radio offset is smaller than the low-frequency one is 136 for $K$ vs. $S/X$, 103 for $X/Ka$ vs. $S/X$, and 99 for $X/Ka$ vs. $K$.
    
    Analysis of \textit{Gaia}/VLBI offsets allowed us to make three conclusions:
     \begin{enumerate}
         \item The hypothesis that \textit{Gaia}-to-$X/Ka$ band offsets are different than \textit{Gaia}-to-$S/X$ band offsets is rejected at the statistical significance level 93\%.
         \item The hypothesis that \textit{Gaia}-to-$K$ band offsets are different than \textit{Gaia}-to-$S/X$ band offsets is rejected at the statistical significance level 25\%.
         \item The hypothesis that for a sample of 228 sources with SI $\geq 3$ \textit{Gaia}-to-$K$ band offsets and \textit{Gaia}-to-$S/X$ band offsets are systematically different is accepted at the statistical significance level 99.9\%.
     \end{enumerate}

\section{Correlation between optical-to-radio offsets and source properties}    \label{sec:r2o-corr}

    In order to understand the origin of optical-to-radio offsets, we investigated the connection between optical-to-radio offsets and source properties such as the source structure index, $G$ magnitude, redshift, and source type.
    We first visually checked the correlation using scatter plots, as shown in Figs.~\ref{fig:rho-vs-si}--\ref{fig:rho-z}.
    Then we used the Nadaraya-Waston estimator \citep{nadaraya1964estimating,watson1964smooth} to characterize empirically the relation between optical-to-radio offsets and source property parameters, as marked by the red dashed lines in these figures.
    The Nadaraya-Waston estimator can be considered as a local average of the response variable weighted by the kernel (we used the Gaussian kernel here) with the benefit of being independent of the binning width choice \citep{2012msma.book.....F}.
    Finally, we performed the non-parametric Spearman's $\rho_S$ rank test on the data points in these plots to check the possible connection between quantities. 
    The correlation coefficient $\rho_S$ and the corresponding $p$-value are also given therein  (The $p$-value roughly indicates the probability of rejecting the null hypothesis, i.e., there exists a correlation) .
    We considered it to be a genuine correlation if the confidence level given in the correlation test is at least 95\%, which means that the $p$-value should be smaller than 0.05.
    The Kendall's $\tau_K$ correlation measure was also computed, which produced consistent results with the Spearman's $\rho_S$ test.
    
\subsection{Source structure}    \label{subsec:r2o-si}

    \begin{figure}[hbtp]
        \centering
        \includegraphics[width=70mm]{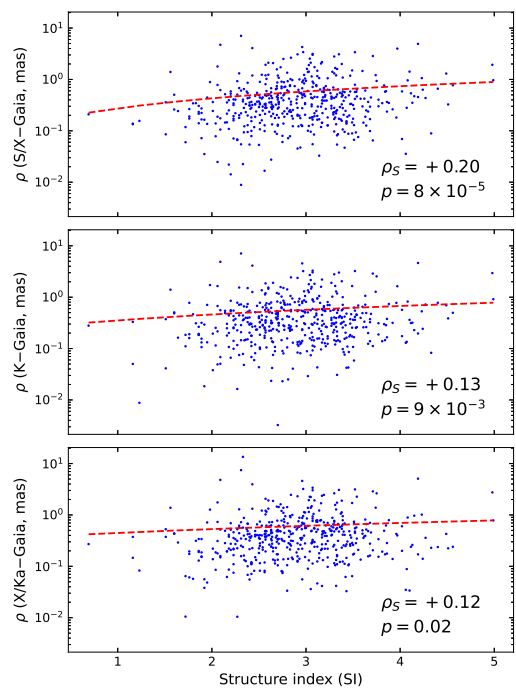}
        \caption[]{\label{fig:rho-vs-si}
            Optical-to-radio offsets at $S/X$, $K$, and $X/Ka$ band as a function of source structure index for 482 sources.
            The red line indicates the Nadaraya-Waston estimator for the relation between the optical-to-radio offsets and the structure index.}
    \end{figure}

    The structure index at $X$ band is available for 482 sources (381 directly from BVID and 101 calculated based on Astrogeo images) out of the 512 sources in our sample, most of which (90\%) fall in the range of 2 to 4.
    A slightly increasing trend is observed at all three bands in Fig.~\ref{fig:rho-vs-si}, which is more evident at $S/X$ band.
    The Spearman test suggests a correlation of $+0.1$ to $+0.2$ between optical-to-radio distance and structure index with confidence level of 98\% or higher (Table~\ref{tab:corr_test}), hence indicating a genuine correlation.
    We also used all the common sources between the \textit{Gaia}-CRF3 and the ICRF3 catalogs at each band to perform this test and found similar correlation. 

\subsection{Magnitude}    \label{subsec:r2o-mag}


    \begin{figure*}[hbtp]
        \centering
        \includegraphics[width=70mm]{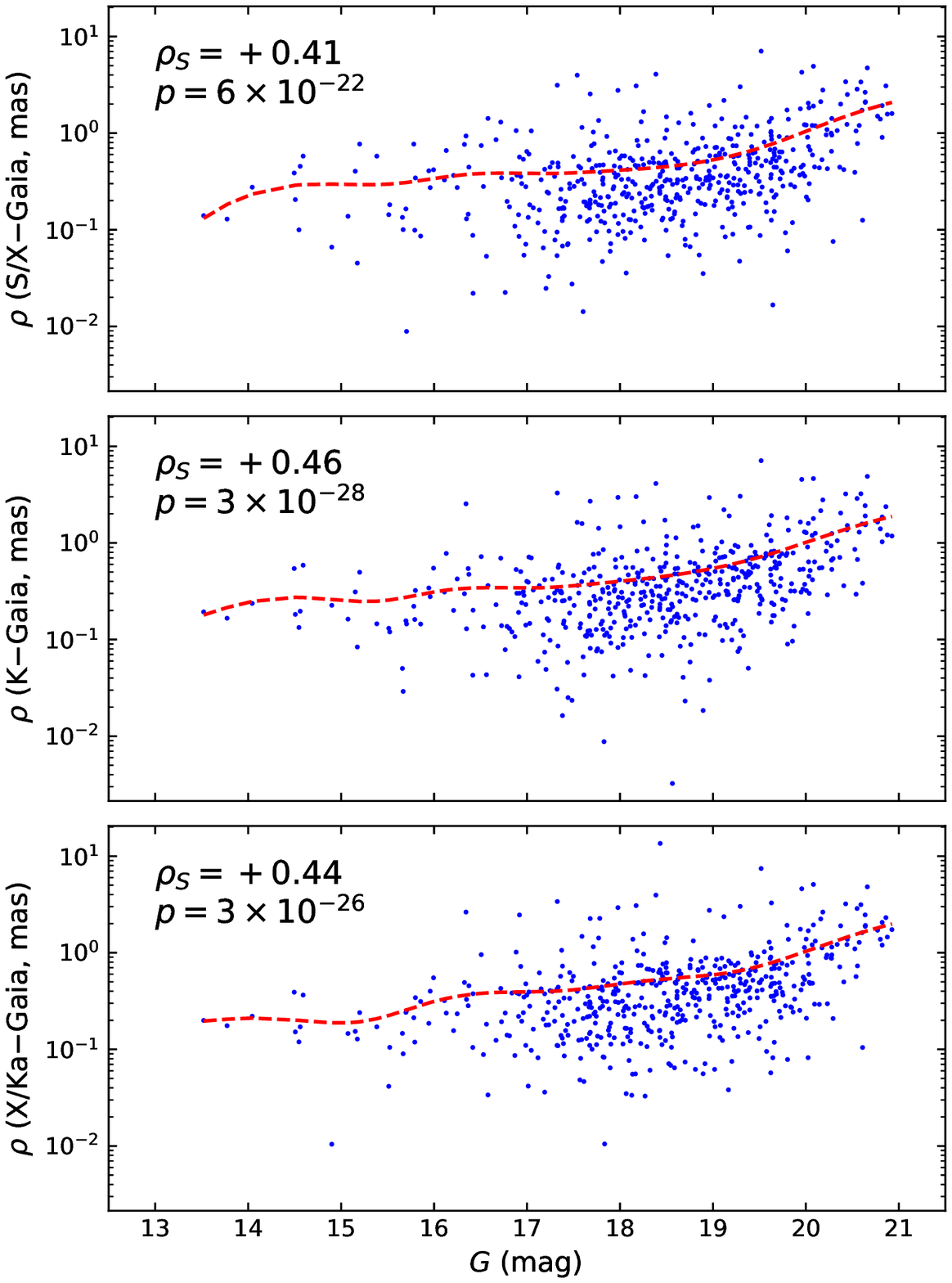}
        \includegraphics[width=70mm]{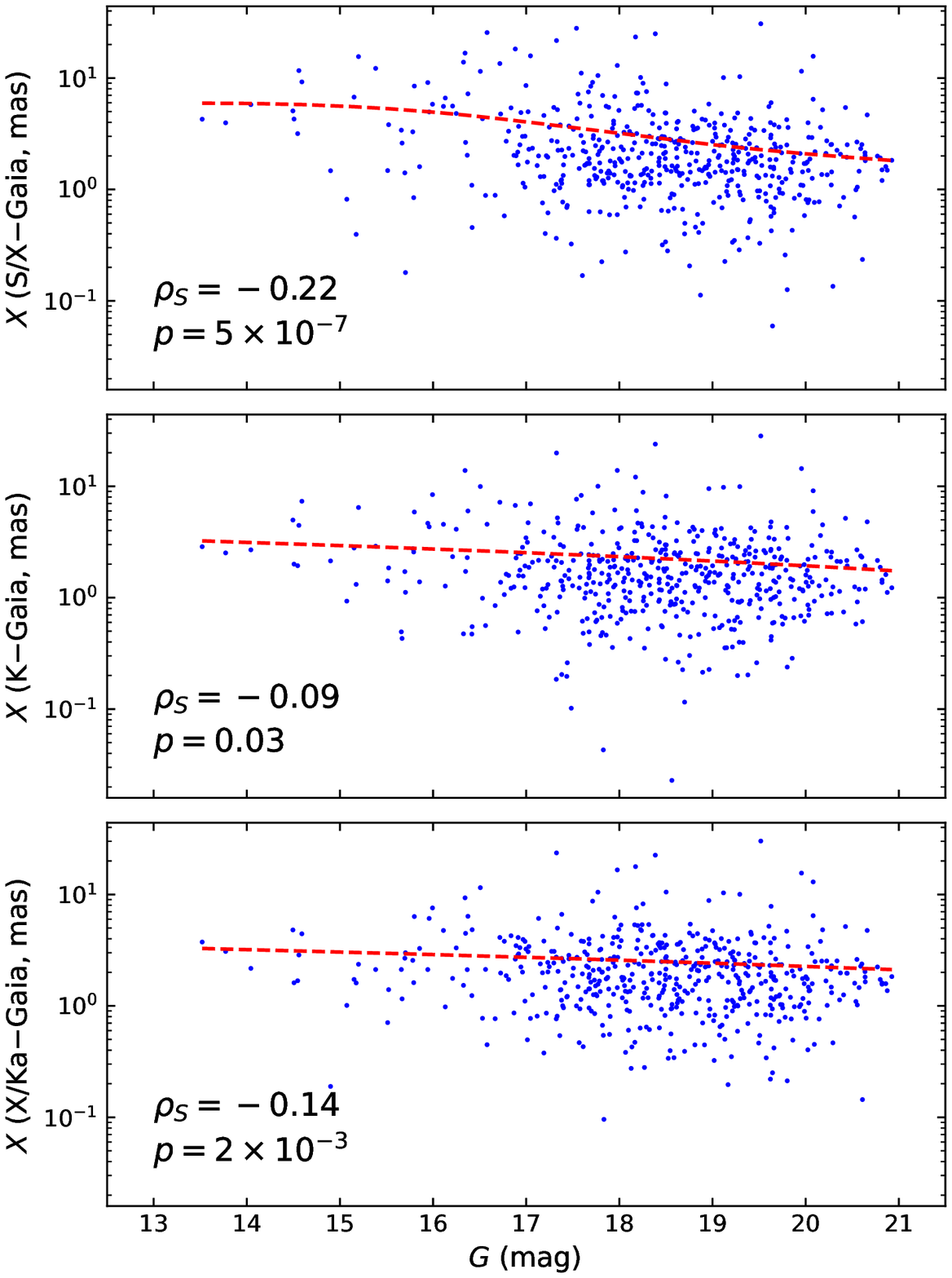}
        \caption[]{\label{fig:rho-g-mag}
            Optical-to-radio offsets (\textit{left}) and normalized separations (\textit{right}) at $S/X$, $K$, and $X/Ka$ band as a function of the \textit{Gaia} $G$ magnitude for the 512 sources in our sample.
            The red line indicates the Nadaraya-Waston estimator for the relation between the optical-to-radio offsets and the $G$ magnitude.}
    \end{figure*}

    The source magnitude is available for all 512 sources directly from the \textit{Gaia} data.
    Figure~\ref{fig:rho-g-mag} demonstrates an obvious dependency of optical-to-radio offsets on the \textit{Gaia} $G$ magnitude: the optical-to-radio offsets increase towards the faint end for all three radio bands.
    This correlation is also seen if considering all common sources between each ICRF3 individual catalog and \textit{Gaia}-CRF3 catalog.
    The correlation test further suggests a quite strong correlation (correlation coefficient > $+0.4$) at all three radio bands with a confidence level higher than 99\% (Table~\ref{tab:corr_test}).
    Considering that the \textit{Gaia} position uncertainty increases with the magnitude, we also checked the correlation between the normalized separation $X$ and the $G$ magnitude.
    As shown in the right panel of Fig.~\ref{fig:rho-g-mag}, $X$ generally decreases with the $G$ magnitude at $S/X$ band and $K$ band; this decreasing tendency is more pronounced at $S/X$ band.
    The negative correlation between normalized separations and the $G$ magnitude is further supported by the results of the correlation tests (reported in Table~\ref{tab:corr_test}).
    As a check, we repeated these analyses on \textit{Gaia}'s BP (blue photometer covering wavelength range 330--680~nm) and RP (red photometer covering wavelength range 640--1050~nm) magnitude and obtained similar results.
    
\subsection{Redshift}    \label{subsec:r2o-z}


    \begin{figure}[hbtp]
        \centering
        \includegraphics[width=70mm]{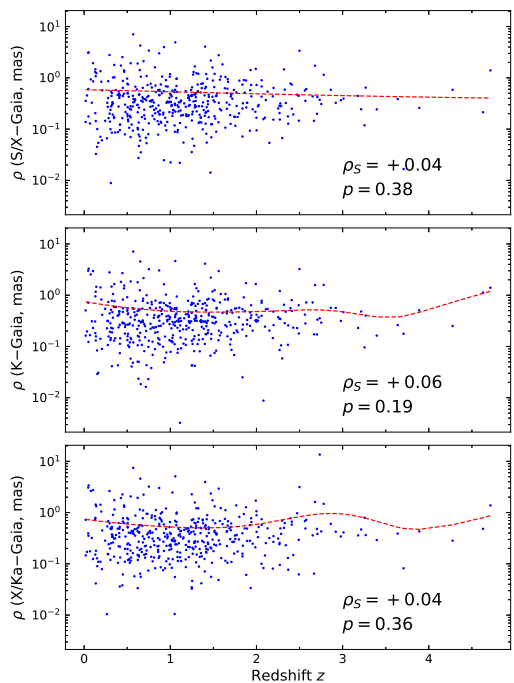}
        \caption[]{\label{fig:rho-z}
            Optical-to-radio offsets at $S/X$, $K$, and $X/Ka$ band as a function of the redshift $z$ for 456 sources having a redshift measurement in the LQAC-5 catalog.
            The red line indicates the Nadaraya-Waston estimator of the relation between the optical-to-radio offsets and the redshift.
        }
    \end{figure}

    We found redshift measurements in the LQAC-5 catalog for 456 sources, the value of which is between 0.5 and 1.5 for more than half of the sources.
    For most sources (80\%), the redshift is smaller than 2.
    As shown in Fig.~\ref{fig:rho-z}, there is no indication of a dependency of the optical-to-radio distance on the redshift, which is also supported by results of the correlation tests (see Table~\ref{tab:corr_test}).

\subsection{Source type}    \label{subsec:r2o-type}
	
	We looked for the source type of these sources in the Optical Characteristics of Astrometric Radio Sources catalog \citep[OCARS;][]{2018ApJS..239...20M}, and found that 355 of them were classified as quasars (labelled as ``AQ'' in the OCARS catalog), while 104 were classified as BL Lac objects (labelled as ``AL''), and 34 as Seyfert 1 galaxies (labelled as ``A1'').
	Then we compared the distribution of the optical-to-radio offsets for the subsets of sources comprised in each class but did not find any significant difference between the three subsets, meaning that the optical-to-radio offset does not depend on the source type.

    \begin{table}
        \centering
        \caption{\label{tab:corr_test}
            Spearman correlation coefficients between the optical-to-radio offsets and source properties for the sample of 512 sources used in this work.
        }
        \begin{tabular}{ccccc}
        \hline \noalign{\smallskip}
        &Nb Source &$S/X$ &$K$  &$X/Ka$  \\
        \noalign{\smallskip}
        \hline
        \noalign{\smallskip}
        $\rho$ vs. SI 
        &482
        &$+0.20$  &$+0.13$  &$+0.12$  \\
        &
        &$\mathit{8 \times 10^{-5}}$  
        &$\mathit{9 \times 10^{-3}}$  
        &$\mathit{0.02}$   \\
        $\rho$ vs. $G$  
        &512
        &$+0.41$  &$+0.46$  &$+0.44$  \\
        &
        &$\mathit{6 \times 10^{-22}}$  
        &$\mathit{3 \times 10^{-28}}$  
        &$\mathit{3 \times 10^{-26}}$  \\
        $X$ vs. $G$ 
        &512
        &$-0.22$  &$-0.09$  &$-0.14$  \\
        &
        &$\mathit{5 \times 10^{-7}}$  
        &$\mathit{0.03}$  
        &$\mathit{2 \times 10^{-3}}$  \\
        $\rho$ vs. $z$  
        &456
        &$+0.04$  &$+0.06$  &$+0.04$  \\
        &
        &\textit{0.38}  &\textit{0.19}  &\textit{0.36}  \\
      \hline \noalign{\smallskip}
      \end{tabular}
      \tablefoot{For each of the comparisons, the first row in the table indicates the correlation coefficient $\rho_s$ while the second row (in italics) provides the corresponding double-sided $p$-value. 
      The $p$-value indicates the probability of rejecting the null hypothesis, i.e., existence of correlation.}
  \end{table}


\section{Discussion} \label{sec:discussion}


\subsection{Cause of optical-to-radio distance} \label{subsec:cause-of-VG}
    \citet{2017MNRAS.471.3775P} summarize several causes for non-coincidence of the emission centers measured by VLBI and \textit{Gaia}.
    These include:\\
    (1) large uncertainties in the \textit{Gaia} or/and VLBI positions;\\
    (2) optical structure (jet) at mas-scale; \\
    (3) optical position shift due to luminous host galaxy or asymmetric structure;\\
    (4) radio source structure and core-shift effect;\\
    (5) gravitational lensing and dual AGNs.\\
    We checked these items based on our results except the second one that has been studied deeply in \citet{2017A&A...598L...1K,2017MNRAS.467L..71P,2017MNRAS.471.3775P,2019MNRAS.482.3023P,2019ApJ...871..143P,2020MNRAS.493L..54K}, and the fifth one which happens only in very few cases.

    Our previous work \citep[][]{2020A&A...634A..28L} indicates that global deformations between optical and radio catalogs may bias optical-to-radio offset studies.
    Based on the sample of 512 sources common to the \textit{Gaia} EDR3 catalog investigated here, we reported a declination bias of $\sim-0.3~{\rm mas}$ in the $X/Ka$ band positions with respect to the \textit{Gaia}-EDR3 positions, similar to that found in the comparison to the \textit{Gaia}-CRF2 positions.
    The existence of such a declination bias would increase the \textit{Gaia}-to-$X/Ka$ band offset by about 0.1--0.2 mas.
    For this reason, large-scale systematics in the $X/Ka$ band frame should be paid attention to.
    On the other hand, only marginal effects are found for the \textit{Gaia}-to-$S/X$ band offsets or \textit{Gaia}-to-$K$ band offsets because relative deformations between those catalogs are much reduced.

    The \textit{Gaia} position uncertainties grow with an increase of optic $G$ magnitude \citep{2016A&A...595A...5M,2018A&A...616A..14G}, while there is no such a dependency for VLBI.
    We observed that the optical-to-radio offsets increase with the $G$ magnitude, especially in the range $17<G<21$ (Fig.~\ref{fig:rho-g-mag}).
    This correlation is further supported by the correlation tests.
    Meanwhile, we also noted that the normalized separations between VLBI and \textit{Gaia} positions generally decrease with the $G$ magnitude.
    A likely explanation is that large optical-to-radio offsets in faint sources are at least partly due to large \textit{Gaia} position uncertainties.
    Those large (but not significant) optical-to-radio offsets thus most probably do not have an astrophysical origin.

    The radio source structure seen at the lower frequencies might shift the observed position at $S/X$ band while the $K$ and $X/Ka$ band positions would be less affected.
    Had most of the sources had significant structure, the optical-to-radio distance would have statistically decreased at higher frequencies.
    However, we do not observe such a tendency nor find any statistical evidence in this sense based on our data.
    Instead, the optical-to-radio offsets in the different bands are roughly at the same level, with the deviation between the bands of less than 0.5~mas for most sources.
    On the other hand, if we limit the sample to the sources with SI~>~3, that is, sources with extended structures, we do find that the optical-to-radio offset is statistically smaller at $K$ band than that at $S/X$ band.   
    This suggests that large optical-to-radio offsets could be a manifestation of extended source structures, a finding previously reported by \citet{2020A&A...644A.159C} based on examining the structure index for the sources that show significant offsets, as derived from comparing the ICRF3 ($S/X$ band) and \textit{Gaia}-CRF2 positions. 
    As noted above, the $X/Ka$ band position is not found to be closer to the optical position than the $S/X$ or $K$ band positions. 
    However, it is difficult to make firm conclusions on it due to the systematics in the $X/Ka$ band frame.
    
    The correlations between the optical-to-radio distances and structure index derived from the $X$ band VLBI images remain weak compared to those between optical-to-radio distances and the $G$ magnitude (Table~\ref{tab:corr_test}).
    To further test this connection, we also examined the structure index measurement at $K$ band from the BVID, which is available for 224 sources of our sources.
    By following the same procedure as that described in Sect.~\ref{subsec:r2o-si}, we obtained Spearman correlation coefficients of $\sim+0.2$ with a confidence level higher than 95\%, suggesting again the existence of a correlation between the optical-to-radio distance and structure index.
    Since no images at $Ka$ band were made, we could not compute SI at that band.
    Combining the results from correlation tests in the $X$ and $K$ bands, the optical-to-radio distance is found to correlate positively with the structure index, which also supports that large optical-to-radio offsets could be due to the extended source structure.
    
    \citet{2019ApJS..242....5X} proposed a quantity, namely the closure amplitude root-mean-square (CARMS), for characterizing the compactness of the sources based on closure observables.
    We also studied the correlation between this quantity and optical-to-radio distances using the sample of 464 sources for which it is available. 
    However, we found a lack of connection (correlation coefficient smaller than $+0.1$ with a $p$-value of about 0.4).
    Since the CARMS was supposed to be correlated with the structure index \citep{2019ApJS..242....5X} and \citet{2021A&A...647A.189X} reported the existence of a connection between the \textit{Gaia}-VLBI and the CARMS, it is surprising that the correlation tests lead to different results.
    We have no explanation yet in this regard and leave it for a future investigation.
    
    We noted that the LQAC-5 catalog provides optical morphological indices derived from the $B$, $R$, and $IR$ DSS (Digital Sky Survey) images, which could be used to infer the existence of a host galaxy.
    We also computed correlations between morphological indices and optical-to-radio offsets, but no statistically-significant correlation was found. 
    Since there are only a few cases with morphological indices larger than one, the influence of the host galaxy on the \textit{Gaia} position, if it exists, is not predominant for the bulk of our sample.
    It is worth noting that these morphological indices were determined from optical imaging with a resolution at the arc-second level. 
    Thus, they could not be used to probe the mas-scale optical jet.
    Morphological indices based on high-resolution images might be useful to probe the existence of the mas-scale optical jets suggested by \citet{2017MNRAS.467L..71P}.

    \citet{2017ApJ...835L..30M} found that the optical-to-radio distance at $S/X$ band decreases rapidly with the redshift $z$ at $z<0.5$.
    They infer that the shift of the \textit{Gaia} positions due to extended optical structures explains partly this finding, noting that the corresponding effect becomes smaller for distant sources.
    Our sample does not support such explanation.
    However, the number of sources located at $z<0.5$ in our sample is limited, preventing us from drawing any firm conclusion about it.
    
    Previous studies based on ground-based optical observations also investigated the correlation between the optical-to-radio offset and source properties.
    However, these authors obtained inconsistent results.
    \citet{2011A&A...532A.115C,2014AJ....147...95Z} reported that the optical-to-radio offset increases with the structure index, while \citet{2013MNRAS.430.2797A} did not detect any dependency.
    \citet{2014AJ....147...95Z} found a negative correlation between the optical-to-radio offset and the redshift but this relation is not found in \citet{2013A&A...553A..13O}.
    To further investigate this matter, we took their catalogs and cross-matched with the \textit{Gaia} EDR3.
    Then, if we use the \textit{Gaia} positions instead, one easily finds that the optical-to-radio offsets as large as 20-40~mas reported in \cite{2013MNRAS.430.2797A} and \cite{2014AJ....147...95Z} largely vanish, turning to values mostly less than 1~mas with \textit{Gaia}.
    \citet{2016A&A...595A...5M} report similar results when comparing the optical-to-radio offsets from the \textit{Gaia} DR1 with those from \cite{2014AJ....147...95Z}.
    Most likely, these large optical-to-radio offsets only reveal artifacts due to unaccounted errors in the ground-based optical astrometry.

\subsection{Implication for the alignment of the optical and radio frames} \label{subsec:sys-effect}

    One reason to extend the ICRF to the $K$ and $X/Ka$ band is that VLBI observations at high frequencies are supposed to suffer less from radio source structure.
    Another reason is that the VLBI core at high frequency may be located closer to the optical-emitting center, making the $K$ and $X/Ka$ band frames perhaps more suitable for the radio-optical frame alignment \citep[e.g.,][]{2002ivsg.conf..350J}.
    We found evidence to support that the $K$-to-\textit{Gaia} position offset is statistically smaller than the $S/X$-to-\textit{Gaia} offset for one third of the sources in our sample, namely those that show extended structures.
    However, when considering all common sources, the $K$-to-\textit{Gaia} and $X/Ka$-to-\textit{Gaia} offsets are found to be no smaller than the $S/X$-to-\textit{Gaia} one.
    As shown here and in our previous study \citep{2020A&A...634A..28L}, the overall agreement between the ICRF3 $K$ and $X/Ka$ band catalogs and the \textit{Gaia}-CRF3 solution remains no higher than for the ICRF3 $S/X$ band catalog.
    Besides, the sample of sources in common with \textit{Gaia} in the ICRF3 $K$ and $X/Ka$ band catalogs is also much smaller.
    Those factors suggest that so far the $K$ and $X/Ka$ band frames would not be a better choice for the alignment of the optical and radio frames than the $S/X$ band frame.

    Previous authors \citep{2008A&A...490..403B,2012MmSAI..83..952M} suggested to base the optical-to-radio frame alignment on sources that have no extended radio structures and are optically-bright (e.g., with magnitude < 18), in other words, with a high astrometric accuracy on both the VLBI and \textit{Gaia} sides.
    In this respect, we detected a connection between the optical-to-radio offset and the source structure index, justifying that the SI is a good indicator for the relevant source selection.
    We also found a strong negative correlation between the optical-to-radio offsets and the \textit{Gaia} $G$ magnitude at all three radio bands.
    This correlation is likely caused by the increased \textit{Gaia} positional uncertainty at faint end.
    As a result, the large optical-to-radio offsets for faint sources are often not statistically significant when considering their uncertainties, meaning that these offsets most probably are not linked to source properties such as extended radio structures or core-shift effects \citep[e.g.,][]{2008A&A...483..759K,2009A&A...505L...1P}.
    In this case, including faint sources in the sample of sources used for the alignment should only add random noise.
    On the other hand, it will greatly enlarge the sample of common sources between radio and optical.
    \citet{2020A&A...634A..28L} found that the accuracy of the \textit{Gaia}-CRF2 did not degrade when increasing the magnitude limit of the sample.
    Hence, the objects for the alignment between the \textit{Gaia}-CRF and the ICRF may not have to be optically bright.


\section{Conclusions} \label{sec:conclusions}

    For the first time, we achieved a multifrequency comparison of extragalactic source positions targeted to study the frequency dependence of those positions.
    To this end, we used a sample of 512 extragalactic sources with positions in the radio $S/X$, $K$, and $X/Ka$ radio bands available in ICRF3 and with the optical positions known from the \textit{Gaia} EDR3 catalog.
    Our main findings are the following:
    \begin{enumerate}
    \item  The existence of large-scale systematics in the ICRF3 $X/Ka$ band catalog distorts the optical-radio vector. 
    Unaccounted, these systematic errors increase the optical-radio distance by 0.1--0.2~mas in average.
    \item The differences between \textit{Gaia}-to-$S/X$ band, \textit{Gaia}-to-$K$ band, and \textit{Gaia}-to-$X/Ka$ band position offsets is statistically insignificant for the entire data set of 512 sources we investigated.
    \item However, the \textit{Gaia}-to-$K$ band distance is shorter than the \textit{Gaia}-to-$S/X$ band one for a subsample of 228 sources with structure index > 3. 
    This result is statistically significant at a 99\% level.
    \item The correlation coefficient between optical-radio distance and the structure index is in a range of 0.12--0.18. 
     This correlation is statistically significant at 98\% level.
    \item The optical-radio distance increases with the \textit{Gaia} $G$ magnitude but shows a generally decreasing trend when normalized by the uncertainty.
    \end{enumerate}
    Based on our results, the ICRF3 $S/X$ band frame remains the preferred choice for aligning the \textit{Gaia}-CRF onto the ICRF due to (i) smaller systematics, (ii) more sources in common to \textit{Gaia} catalogs, and (iii) the fact that the \textit{Gaia}-to-$X/Ka$ and \textit{Gaia}-to-$K$ offset is not statistically smaller than the \textit{Gaia}-to-$S/X$ one when considering all sources in our data.
    The efforts to improve the $K$ and $X/Ka$ band frames should be continued, which should help to further assess the dependence of optical-to-radio distance on source structure, magnitude, and redshift.

\begin{acknowledgements}
    We sincerely thank the anonymous referee for the constructive comments and useful suggestions, which improve the work a lot.
    N.L. and Z.Z. are supported by the National Natural Science Foundation of China (NSFC) under grant No. 11833004.
    N.L. is also supported by the Fundamental Research Funds for the Central Universities of China under grant No. 14380042.
    This work has made use of data from the European Space Agency (ESA) mission {\it Gaia} (\url{https://www.cosmos.esa.int/gaia}), processed by the {\it Gaia} Data Processing and Analysis Consortium (DPAC, \url{https://www.cosmos.esa.int/web/gaia/dpac/consortium}).
    Funding for the DPAC has been provided by national institutions, in particular the institutions participating in the {\it Gaia} Multilateral Agreement.
    This research has made use of material from the Bordeaux VLBI Image Database (BVID).
    This database can be reached at \url{http://bvid.astrophy.u-bordeaux.fr/}.
    Some radio images were retrieved from the Astrogeo VLBI FITS image database (\url{http://astrogeo.org/vlbi_images/}).
    This research had also made use of Astropy\footnote{\href{http://www.astropy.org}{http://www.astropy.org}} -- a community-developed core Python package for Astronomy \citep{2018AJ....156..123A}, the Python 2D plotting library Matplotlib \citep{2007CSE.....9...90H}, and TOPCAT \citep{2011ascl.soft01010T}.
    We made much use of NASA's Astrophysics Data System and the VizieR catalogue access tool, CDS, Strasbourg, France (DOI : \href{https://doi.org/10.26093/cds/vizier}{10.26093/cds/vizier}). 
    The original description of the VizieR service was published in \citet{2000A&AS..143...23O}.

\end{acknowledgements}


\bibliographystyle{aa} 
\bibliography{references}          

\begin{thebibliography}{41}
\expandafter\ifx\csname natexlab\endcsname\relax\def\natexlab#1{#1}\fi

\bibitem[{{Aslan} {et~al.}(2010){Aslan}, {Gumerov}, {Jin}, {Khamitov},
  {Maigurova}, {Pinigin}, {Tang}, \& {Wang}}]{2010A&A...510A..10A}
{Aslan}, Z., {Gumerov}, R., {Jin}, W., {et~al.} 2010, \aap, 510, A10

\bibitem[{{Assafin} {et~al.}(2013){Assafin}, {Vieira-Martins}, {Andrei},
  {Camargo}, \& {da Silva Neto}}]{2013MNRAS.430.2797A}
{Assafin}, M., {Vieira-Martins}, R., {Andrei}, A.~H., {Camargo}, J.~I.~B., \&
  {da Silva Neto}, D.~N. 2013, \mnras, 430, 2797

\bibitem[{{Astropy Collaboration} {et~al.}(2018){Astropy Collaboration},
  {Price-Whelan}, {Sip{\H{o}}cz}, {G{\"u}nther}, {Lim}, {Crawford}, {Conseil},
  {Shupe}, {Craig}, {Dencheva}, {Ginsburg}, {VanderPlas}, {Bradley},
  {P{\'e}rez-Su{\'a}rez}, {de Val-Borro}, {Aldcroft}, {Cruz}, {Robitaille},
  {Tollerud}, {Ardelean}, {Babej}, {Bach}, {Bachetti}, {Bakanov}, {Bamford},
  {Barentsen}, {Barmby}, {Baumbach}, {Berry}, {Biscani}, {Boquien}, {Bostroem},
  {Bouma}, {Brammer}, {Bray}, {Breytenbach}, {Buddelmeijer}, {Burke},
  {Calderone}, {Cano Rodr{\'\i}guez}, {Cara}, {Cardoso}, {Cheedella}, {Copin},
  {Corrales}, {Crichton}, {D'Avella}, {Deil}, {Depagne}, {Dietrich}, {Donath},
  {Droettboom}, {Earl}, {Erben}, {Fabbro}, {Ferreira}, {Finethy}, {Fox},
  {Garrison}, {Gibbons}, {Goldstein}, {Gommers}, {Greco}, {Greenfield},
  {Groener}, {Grollier}, {Hagen}, {Hirst}, {Homeier}, {Horton}, {Hosseinzadeh},
  {Hu}, {Hunkeler}, {Ivezi{\'c}}, {Jain}, {Jenness}, {Kanarek}, {Kendrew},
  {Kern}, {Kerzendorf}, {Khvalko}, {King}, {Kirkby}, {Kulkarni}, {Kumar},
  {Lee}, {Lenz}, {Littlefair}, {Ma}, {Macleod}, {Mastropietro}, {McCully},
  {Montagnac}, {Morris}, {Mueller}, {Mumford}, {Muna}, {Murphy}, {Nelson},
  {Nguyen}, {Ninan}, {N{\"o}the}, {Ogaz}, {Oh}, {Parejko}, {Parley}, {Pascual},
  {Patil}, {Patil}, {Plunkett}, {Prochaska}, {Rastogi}, {Reddy Janga},
  {Sabater}, {Sakurikar}, {Seifert}, {Sherbert}, {Sherwood-Taylor}, {Shih},
  {Sick}, {Silbiger}, {Singanamalla}, {Singer}, {Sladen}, {Sooley},
  {Sornarajah}, {Streicher}, {Teuben}, {Thomas}, {Tremblay}, {Turner},
  {Terr{\'o}n}, {van Kerkwijk}, {de la Vega}, {Watkins}, {Weaver}, {Whitmore},
  {Woillez}, {Zabalza}, \& {Astropy Contributors}}]{2018AJ....156..123A}
{Astropy Collaboration}, {Price-Whelan}, A.~M., {Sip{\H{o}}cz}, B.~M., {et~al.}
  2018, \aj, 156, 123

\bibitem[{{Bourda} {et~al.}(2008){Bourda}, {Charlot}, \& {Le
  Campion}}]{2008A&A...490..403B}
{Bourda}, G., {Charlot}, P., \& {Le Campion}, J.~F. 2008, \aap, 490, 403

\bibitem[{{Camargo} {et~al.}(2011){Camargo}, {Andrei}, {Assafin},
  {Vieira-Martins}, \& {da Silva Neto}}]{2011A&A...532A.115C}
{Camargo}, J.~I.~B., {Andrei}, A.~H., {Assafin}, M., {Vieira-Martins}, R., \&
  {da Silva Neto}, D.~N. 2011, \aap, 532, A115

\bibitem[{{Charlot} {et~al.}(2010){Charlot}, {Boboltz}, {Fey}, {Fomalont},
  {Geldzahler}, {Gordon}, {Jacobs}, {Lanyi}, {Ma}, {Naudet}, {Romney},
  {Sovers}, \& {Zhang}}]{2010AJ....139.1713C}
{Charlot}, P., {Boboltz}, D.~A., {Fey}, A.~L., {et~al.} 2010, \aj, 139, 1713

\bibitem[{{Charlot} {et~al.}(2020){Charlot}, {Jacobs}, {Gordon}, {Lambert}, {de
  Witt}, {B{\"o}hm}, {Fey}, {Heinkelmann}, {Skurikhina}, {Titov}, {Arias},
  {Bolotin}, {Bourda}, {Ma}, {Malkin}, {Nothnagel}, {Mayer}, {MacMillan},
  {Nilsson}, \& {Gaume}}]{2020A&A...644A.159C}
{Charlot}, P., {Jacobs}, C.~S., {Gordon}, D., {et~al.} 2020, \aap, 644, A159

\bibitem[{{Feigelson} \& {Babu}(2012)}]{2012msma.book.....F}
{Feigelson}, E.~D. \& {Babu}, G.~J. 2012, {Modern Statistical Methods for
  Astronomy}

\bibitem[{{Fey} \& {Charlot}(1997)}]{1997ApJS..111...95F}
{Fey}, A.~L. \& {Charlot}, P. 1997, \apjs, 111, 95

\bibitem[{{Fey} {et~al.}(2015){Fey}, {Gordon}, {Jacobs}, {Ma}, {Gaume},
  {Arias}, {Bianco}, {Boboltz}, {B{\"o}ckmann}, {Bolotin}, {Charlot},
  {Collioud}, {Engelhardt}, {Gipson}, {Gontier}, {Heinkelmann}, {Kurdubov},
  {Lambert}, {Lytvyn}, {MacMillan}, {Malkin}, {Nothnagel}, {Ojha},
  {Skurikhina}, {Sokolova}, {Souchay}, {Sovers}, {Tesmer}, {Titov}, {Wang}, \&
  {Zharov}}]{2015AJ....150...58F}
{Fey}, A.~L., {Gordon}, D., {Jacobs}, C.~S., {et~al.} 2015, \aj, 150, 58

\bibitem[{{Frouard} {et~al.}(2018){Frouard}, {Johnson}, {Fey}, {Makarov}, \&
  {Dorland}}]{2018AJ....155..229F}
{Frouard}, J., {Johnson}, M.~C., {Fey}, A., {Makarov}, V.~V., \& {Dorland},
  B.~N. 2018, \aj, 155, 229

\bibitem[{{Gaia Collaboration} {et~al.}(2021){Gaia Collaboration}, {Brown},
  {Vallenari}, {Prusti}, {de Bruijne}, {Babusiaux}, {Biermann}, {Creevey},
  {Evans}, {Eyer}, {Hutton}, {Jansen}, {Jordi}, {Klioner}, {Lammers},
  {Lindegren}, {Luri}, {Mignard}, {Panem}, {Pourbaix}, {Randich}, {Sartoretti},
  {Soubiran}, {Walton}, {Arenou}, {Bailer-Jones}, {Bastian}, {Cropper},
  {Drimmel}, {Katz}, {Lattanzi}, {van Leeuwen}, {Bakker}, {Cacciari},
  {Casta{\~n}eda}, {De Angeli}, {Ducourant}, {Fabricius}, {Fouesneau},
  {Fr{\'e}mat}, {Guerra}, {Guerrier}, {Guiraud}, {Jean-Antoine Piccolo},
  {Masana}, {Messineo}, {Mowlavi}, {Nicolas}, {Nienartowicz}, {Pailler},
  {Panuzzo}, {Riclet}, {Roux}, {Seabroke}, {Sordo}, {Tanga}, {Th{\'e}venin},
  {Gracia-Abril}, {Portell}, {Teyssier}, {Altmann}, {Andrae}, {Bellas-Velidis},
  {Benson}, {Berthier}, {Blomme}, {Brugaletta}, {Burgess}, {Busso}, {Carry},
  {Cellino}, {Cheek}, {Clementini}, {Damerdji}, {Davidson}, {Delchambre},
  {Dell'Oro}, {Fern{\'a}ndez-Hern{\'a}ndez}, {Galluccio}, {Garc{\'\i}a-Lario},
  {Garcia-Reinaldos}, {Gonz{\'a}lez-N{\'u}{\~n}ez}, {Gosset}, {Haigron},
  {Halbwachs}, {Hambly}, {Harrison}, {Hatzidimitriou}, {Heiter},
  {Hern{\'a}ndez}, {Hestroffer}, {Hodgkin}, {Holl}, {Jan{\ss}en}, {Jevardat de
  Fombelle}, {Jordan}, {Krone-Martins}, {Lanzafame}, {L{\"o}ffler}, {Lorca},
  {Manteiga}, {Marchal}, {Marrese}, {Moitinho}, {Mora}, {Muinonen}, {Osborne},
  {Pancino}, {Pauwels}, {Petit}, {Recio-Blanco}, {Richards}, {Riello},
  {Rimoldini}, {Robin}, {Roegiers}, {Rybizki}, {Sarro}, {Siopis}, {Smith},
  {Sozzetti}, {Ulla}, {Utrilla}, {van Leeuwen}, {van Reeven}, {Abbas}, {Abreu
  Aramburu}, {Accart}, {Aerts}, {Aguado}, {Ajaj}, {Altavilla}, {{\'A}lvarez},
  {{\'A}lvarez Cid-Fuentes}, {Alves}, {Anderson}, {Anglada Varela}, {Antoja},
  {Audard}, {Baines}, {Baker}, {Balaguer-N{\'u}{\~n}ez}, {Balbinot}, {Balog},
  {Barache}, {Barbato}, {Barros}, {Barstow}, {Bartolom{\'e}}, {Bassilana},
  {Bauchet}, {Baudesson-Stella}, {Becciani}, {Bellazzini}, {Bernet}, {Bertone},
  {Bianchi}, {Blanco-Cuaresma}, {Boch}, {Bombrun}, {Bossini}, {Bouquillon},
  {Bragaglia}, {Bramante}, {Breedt}, {Bressan}, {Brouillet}, {Bucciarelli},
  {Burlacu}, {Busonero}, {Butkevich}, {Buzzi}, {Caffau}, {Cancelliere},
  {C{\'a}novas}, {Cantat-Gaudin}, {Carballo}, {Carlucci}, {Carnerero},
  {Carrasco}, {Casamiquela}, {Castellani}, {Castro-Ginard}, {Castro Sampol},
  {Chaoul}, {Charlot}, {Chemin}, {Chiavassa}, {Cioni}, {Comoretto}, {Cooper},
  {Cornez}, {Cowell}, {Crifo}, {Crosta}, {Crowley}, {Dafonte}, {Dapergolas},
  {David}, {David}, {de Laverny}, {De Luise}, {De March}, {De Ridder}, {de
  Souza}, {de Teodoro}, {de Torres}, {del Peloso}, {del Pozo}, {Delbo},
  {Delgado}, {Delgado}, {Delisle}, {Di Matteo}, {Diakite}, {Diener},
  {Distefano}, {Dolding}, {Eappachen}, {Edvardsson}, {Enke}, {Esquej}, {Fabre},
  {Fabrizio}, {Faigler}, {Fedorets}, {Fernique}, {Fienga}, {Figueras},
  {Fouron}, {Fragkoudi}, {Fraile}, {Franke}, {Gai}, {Garabato},
  {Garcia-Gutierrez}, {Garc{\'\i}a-Torres}, {Garofalo}, {Gavras}, {Gerlach},
  {Geyer}, {Giacobbe}, {Gilmore}, {Girona}, {Giuffrida}, {Gomel}, {Gomez},
  {Gonzalez-Santamaria}, {Gonz{\'a}lez-Vidal}, {Granvik},
  {Guti{\'e}rrez-S{\'a}nchez}, {Guy}, {Hauser}, {Haywood}, {Helmi}, {Hidalgo},
  {Hilger}, {H{\l}adczuk}, {Hobbs}, {Holland}, {Huckle}, {Jasniewicz},
  {Jonker}, {Juaristi Campillo}, {Julbe}, {Karbevska}, {Kervella}, {Khanna},
  {Kochoska}, {Kontizas}, {Kordopatis}, {Korn}, {Kostrzewa-Rutkowska},
  {Kruszy{\'n}ska}, {Lambert}, {Lanza}, {Lasne}, {Le Campion}, {Le Fustec},
  {Lebreton}, {Lebzelter}, {Leccia}, {Leclerc}, {Lecoeur-Taibi}, {Liao},
  {Licata}, {Lindstr{\o}m}, {Lister}, {Livanou}, {Lobel}, {Madrero Pardo},
  {Managau}, {Mann}, {Marchant}, {Marconi}, {Marcos Santos}, {Marinoni},
  {Marocco}, {Marshall}, {Martin Polo}, {Mart{\'\i}n-Fleitas}, {Masip},
  {Massari}, {Mastrobuono-Battisti}, {Mazeh}, {McMillan}, {Messina},
  {Michalik}, {Millar}, {Mints}, {Molina}, {Molinaro}, {Moln{\'a}r},
  {Montegriffo}, {Mor}, {Morbidelli}, {Morel}, {Morris}, {Mulone}, {Munoz},
  {Muraveva}, {Murphy}, {Musella}, {Noval}, {Ord{\'e}novic}, {Orr{\`u}},
  {Osinde}, {Pagani}, {Pagano}, {Palaversa}, {Palicio}, {Panahi}, {Pawlak},
  {Pe{\~n}alosa Esteller}, {Penttil{\"a}}, {Piersimoni}, {Pineau}, {Plachy},
  {Plum}, {Poggio}, {Poretti}, {Poujoulet}, {Pr{\v{s}}a}, {Pulone}, {Racero},
  {Ragaini}, {Rainer}, {Raiteri}, {Rambaux}, {Ramos}, {Ramos-Lerate}, {Re
  Fiorentin}, {Regibo}, {Reyl{\'e}}, {Ripepi}, {Riva}, {Rixon}, {Robichon},
  {Robin}, {Roelens}, {Rohrbasser}, {Romero-G{\'o}mez}, {Rowell}, {Royer},
  {Rybicki}, {Sadowski}, {Sagrist{\`a} Sell{\'e}s}, {Sahlmann}, {Salgado},
  {Salguero}, {Samaras}, {Sanchez Gimenez}, {Sanna}, {Santove{\~n}a},
  {Sarasso}, {Schultheis}, {Sciacca}, {Segol}, {Segovia}, {S{\'e}gransan},
  {Semeux}, {Shahaf}, {Siddiqui}, {Siebert}, {Siltala}, {Slezak}, {Smart},
  {Solano}, {Solitro}, {Souami}, {Souchay}, {Spagna}, {Spoto}, {Steele},
  {Steidelm{\"u}ller}, {Stephenson}, {S{\"u}veges}, {Szabados}, {Szegedi-Elek},
  {Taris}, {Tauran}, {Taylor}, {Teixeira}, {Thuillot}, {Tonello}, {Torra},
  {Torra}, {Turon}, {Unger}, {Vaillant}, {van Dillen}, {Vanel}, {Vecchiato},
  {Viala}, {Vicente}, {Voutsinas}, {Weiler}, {Wevers}, {Wyrzykowski}, {Yoldas},
  {Yvard}, {Zhao}, {Zorec}, {Zucker}, {Zurbach}, \&
  {Zwitter}}]{2021A&A...649A...1G}
{Gaia Collaboration}, {Brown}, A.~G.~A., {Vallenari}, A., {et~al.} 2021, \aap,
  649, A1

\bibitem[{{Gaia Collaboration} {et~al.}(2018){Gaia Collaboration}, {Mignard},
  {Klioner}, {Lindegren}, {Hern{\'a}ndez}, {Bastian}, {Bombrun}, {Hobbs},
  {Lammers}, {Michalik}, {Ramos-Lerate}, {Biermann},
  {Fern{\'a}ndez-Hern{\'a}ndez}, {Geyer}, {Hilger}, {Siddiqui},
  {Steidelm{\"u}ller}, {Babusiaux}, {Barache}, {Lambert}, {Andrei}, {Bourda},
  {Charlot}, {Brown}, {Vallenari}, {Prusti}, {de Bruijne}, {Bailer-Jones},
  {Evans}, {Eyer}, {Jansen}, {Jordi}, {Luri}, {Panem}, {Pourbaix}, {Randich},
  {Sartoretti}, {Soubiran}, {van Leeuwen}, {Walton}, {Arenou}, {Cropper},
  {Drimmel}, {Katz}, {Lattanzi}, {Bakker}, {Cacciari}, {Casta{\~n}eda},
  {Chaoul}, {Cheek}, {De Angeli}, {Fabricius}, {Guerra}, {Holl}, {Masana},
  {Messineo}, {Mowlavi}, {Nienartowicz}, {Panuzzo}, {Portell}, {Riello},
  {Seabroke}, {Tanga}, {Th{\'e}venin}, {Gracia-Abril}, {Comoretto},
  {Garcia-Reinaldos}, {Teyssier}, {Altmann}, {Andrae}, {Audard},
  {Bellas-Velidis}, {Benson}, {Berthier}, {Blomme}, {Burgess}, {Busso},
  {Carry}, {Cellino}, {Clementini}, {Clotet}, {Creevey}, {Davidson}, {De
  Ridder}, {Delchambre}, {Dell'Oro}, {Ducourant}, {Fouesneau}, {Fr{\'e}mat},
  {Galluccio}, {Garc{\'\i}a-Torres}, {Gonz{\'a}lez-N{\'u}{\~n}ez},
  {Gonz{\'a}lez-Vidal}, {Gosset}, {Guy}, {Halbwachs}, {Hambly}, {Harrison},
  {Hestroffer}, {Hodgkin}, {Hutton}, {Jasniewicz}, {Jean-Antoine-Piccolo},
  {Jordan}, {Korn}, {Krone-Martins}, {Lanzafame}, {Lebzelter}, {L{\"o}ffler},
  {Manteiga}, {Marrese}, {Mart{\'\i}n-Fleitas}, {Moitinho}, {Mora}, {Muinonen},
  {Osinde}, {Pancino}, {Pauwels}, {Petit}, {Recio-Blanco}, {Richards},
  {Rimoldini}, {Robin}, {Sarro}, {Siopis}, {Smith}, {Sozzetti}, {S{\"u}veges},
  {Torra}, {van Reeven}, {Abbas}, {Abreu Aramburu}, {Accart}, {Aerts},
  {Altavilla}, {{\'A}lvarez}, {Alvarez}, {Alves}, {Anderson}, {Anglada Varela},
  {Antiche}, {Antoja}, {Arcay}, {Astraatmadja}, {Bach}, {Baker},
  {Balaguer-N{\'u}{\~n}ez}, {Balm}, {Barata}, {Barbato}, {Barblan}, {Barklem},
  {Barrado}, {Barros}, {Barstow}, {Bartholom{\'e} Mu{\~n}oz}, {Bassilana},
  {Becciani}, {Bellazzini}, {Berihuete}, {Bertone}, {Bianchi}, {Bienaym{\'e}},
  {Blanco-Cuaresma}, {Boch}, {Boeche}, {Borrachero}, {Bossini}, {Bouquillon},
  {Bragaglia}, {Bramante}, {Breddels}, {Bressan}, {Brouillet},
  {Br{\"u}semeister}, {Brugaletta}, {Bucciarelli}, {Burlacu}, {Busonero},
  {Butkevich}, {Buzzi}, {Caffau}, {Cancelliere}, {Cannizzaro}, {Cantat-Gaudin},
  {Carballo}, {Carlucci}, {Carrasco}, {Casamiquela}, {Castellani},
  {Castro-Ginard}, {Chemin}, {Chiavassa}, {Cocozza}, {Costigan}, {Cowell},
  {Crifo}, {Crosta}, {Crowley}, {Cuypers}, {Dafonte}, {Damerdji}, {Dapergolas},
  {David}, {David}, {de Laverny}, {De Luise}, {De March}, {de Souza}, {de
  Torres}, {Debosscher}, {del Pozo}, {Delbo}, {Delgado}, {Delgado}, {Diakite},
  {Diener}, {Distefano}, {Dolding}, {Drazinos}, {Dur{\'a}n}, {Edvardsson},
  {Enke}, {Eriksson}, {Esquej}, {Eynard Bontemps}, {Fabre}, {Fabrizio},
  {Faigler}, {Falc{\~a}o}, {Farr{\`a}s Casas}, {Federici}, {Fedorets},
  {Fernique}, {Figueras}, {Filippi}, {Findeisen}, {Fonti}, {Fraile}, {Fraser},
  {Fr{\'e}zouls}, {Gai}, {Galleti}, {Garabato}, {Garc{\'\i}a-Sedano},
  {Garofalo}, {Garralda}, {Gavel}, {Gavras}, {Gerssen}, {Giacobbe}, {Gilmore},
  {Girona}, {Giuffrida}, {Glass}, {Gomes}, {Granvik}, {Gueguen}, {Guerrier},
  {Guiraud}, {Guti{\'e}}, {Haigron}, {Hatzidimitriou}, {Hauser}, {Haywood},
  {Heiter}, {Helmi}, {Heu}, {Hofmann}, {Holland}, {Huckle}, {Hypki}, {Icardi},
  {Jan{\ss}en}, {Jevardat de Fombelle}, {Jonker}, {Juh{\'a}sz}, {Julbe},
  {Karampelas}, {Kewley}, {Klar}, {Kochoska}, {Kohley}, {Kolenberg},
  {Kontizas}, {Kontizas}, {Koposov}, {Kordopatis}, {Kostrzewa-Rutkowska},
  {Koubsky}, {Lanza}, {Lasne}, {Lavigne}, {Le Fustec}, {Le Poncin-Lafitte},
  {Lebreton}, {Leccia}, {Leclerc}, {Lecoeur-Taibi}, {Lenhardt}, {Leroux},
  {Liao}, {Licata}, {Lindstr{\o}m}, {Lister}, {Livanou}, {Lobel}, {L{\'o}pez},
  {Managau}, {Mann}, {Mantelet}, {Marchal}, {Marchant}, {Marconi}, {Marinoni},
  {Marschalk{\'o}}, {Marshall}, {Martino}, {Marton}, {Mary}, {Massari},
  {Matijevi{\v{c}}}, {Mazeh}, {McMillan}, {Messina}, {Millar}, {Molina},
  {Molinaro}, {Moln{\'a}r}, {Montegriffo}, {Mor}, {Morbidelli}, {Morel},
  {Morris}, {Mulone}, {Muraveva}, {Musella}, {Nelemans}, {Nicastro}, {Noval},
  {O'Mullane}, {Ord{\'e}novic}, {Ord{\'o}{\~n}ez-Blanco}, {Osborne}, {Pagani},
  {Pagano}, {Pailler}, {Palacin}, {Palaversa}, {Panahi}, {Pawlak},
  {Piersimoni}, {Pineau}, {Plachy}, {Plum}, {Poggio}, {Poujoulet},
  {Pr{\v{s}}a}, {Pulone}, {Racero}, {Ragaini}, {Rambaux}, {Regibo},
  {Reyl{\'e}}, {Riclet}, {Ripepi}, {Riva}, {Rivard}, {Rixon}, {Roegiers},
  {Roelens}, {Romero-G{\'o}mez}, {Rowell}, {Royer}, {Ruiz-Dern}, {Sadowski},
  {Sagrist{\`a} Sell{\'e}s}, {Sahlmann}, {Salgado}, {Salguero}, {Sanna},
  {Santana-Ros}, {Sarasso}, {Savietto}, {Schultheis}, {Sciacca}, {Segol},
  {Segovia}, {S{\'e}gransan}, {Shih}, {Siltala}, {Silva}, {Smart}, {Smith},
  {Solano}, {Solitro}, {Sordo}, {Soria Nieto}, {Souchay}, {Spagna}, {Spoto},
  {Stampa}, {Steele}, {Stephenson}, {Stoev}, {Suess}, {Surdej}, {Szabados},
  {Szegedi-Elek}, {Tapiador}, {Taris}, {Tauran}, {Taylor}, {Teixeira},
  {Terrett}, {Teyssandier}, {Thuillot}, {Titarenko}, {Torra Clotet}, {Turon},
  {Ulla}, {Utrilla}, {Uzzi}, {Vaillant}, {Valentini}, {Valette}, {van Elteren},
  {Van Hemelryck}, {van Leeuwen}, {Vaschetto}, {Vecchiato}, {Veljanoski},
  {Viala}, {Vicente}, {Vogt}, {von Essen}, {Voss}, {Votruba}, {Voutsinas},
  {Walmsley}, {Weiler}, {Wertz}, {Wevers}, {Wyrzykowski}, {Yoldas},
  {{\v{Z}}erjal}, {Ziaeepour}, {Zorec}, {Zschocke}, {Zucker}, {Zurbach}, \&
  {Zwitter}}]{2018A&A...616A..14G}
{Gaia Collaboration}, {Mignard}, F., {Klioner}, S.~A., {et~al.} 2018, \aap,
  616, A14

\bibitem[{{Gaia Collaboration} {et~al.}(2016){Gaia Collaboration}, {Prusti},
  {de Bruijne}, {Brown}, {Vallenari}, {Babusiaux}, {Bailer-Jones}, {Bastian},
  {Biermann}, {Evans}, {Eyer}, {Jansen}, {Jordi}, {Klioner}, {Lammers},
  {Lindegren}, {Luri}, {Mignard}, {Milligan}, {Panem}, {Poinsignon},
  {Pourbaix}, {Randich}, {Sarri}, {Sartoretti}, {Siddiqui}, {Soubiran},
  {Valette}, {van Leeuwen}, {Walton}, {Aerts}, {Arenou}, {Cropper}, {Drimmel},
  {H{\o}g}, {Katz}, {Lattanzi}, {O'Mullane}, {Grebel}, {Holland}, {Huc},
  {Passot}, {Bramante}, {Cacciari}, {Casta{\~n}eda}, {Chaoul}, {Cheek}, {De
  Angeli}, {Fabricius}, {Guerra}, {Hern{\'a}ndez}, {Jean-Antoine-Piccolo},
  {Masana}, {Messineo}, {Mowlavi}, {Nienartowicz}, {Ord{\'o}{\~n}ez-Blanco},
  {Panuzzo}, {Portell}, {Richards}, {Riello}, {Seabroke}, {Tanga},
  {Th{\'e}venin}, {Torra}, {Els}, {Gracia-Abril}, {Comoretto},
  {Garcia-Reinaldos}, {Lock}, {Mercier}, {Altmann}, {Andrae}, {Astraatmadja},
  {Bellas-Velidis}, {Benson}, {Berthier}, {Blomme}, {Busso}, {Carry},
  {Cellino}, {Clementini}, {Cowell}, {Creevey}, {Cuypers}, {Davidson}, {De
  Ridder}, {de Torres}, {Delchambre}, {Dell'Oro}, {Ducourant}, {Fr{\'e}mat},
  {Garc{\'\i}a-Torres}, {Gosset}, {Halbwachs}, {Hambly}, {Harrison}, {Hauser},
  {Hestroffer}, {Hodgkin}, {Huckle}, {Hutton}, {Jasniewicz}, {Jordan},
  {Kontizas}, {Korn}, {Lanzafame}, {Manteiga}, {Moitinho}, {Muinonen},
  {Osinde}, {Pancino}, {Pauwels}, {Petit}, {Recio-Blanco}, {Robin}, {Sarro},
  {Siopis}, {Smith}, {Smith}, {Sozzetti}, {Thuillot}, {van Reeven}, {Viala},
  {Abbas}, {Abreu Aramburu}, {Accart}, {Aguado}, {Allan}, {Allasia},
  {Altavilla}, {{\'A}lvarez}, {Alves}, {Anderson}, {Andrei}, {Anglada Varela},
  {Antiche}, {Antoja}, {Ant{\'o}n}, {Arcay}, {Atzei}, {Ayache}, {Bach},
  {Baker}, {Balaguer-N{\'u}{\~n}ez}, {Barache}, {Barata}, {Barbier}, {Barblan},
  {Baroni}, {Barrado y Navascu{\'e}s}, {Barros}, {Barstow}, {Becciani},
  {Bellazzini}, {Bellei}, {Bello Garc{\'\i}a}, {Belokurov}, {Bendjoya},
  {Berihuete}, {Bianchi}, {Bienaym{\'e}}, {Billebaud}, {Blagorodnova},
  {Blanco-Cuaresma}, {Boch}, {Bombrun}, {Borrachero}, {Bouquillon}, {Bourda},
  {Bouy}, {Bragaglia}, {Breddels}, {Brouillet}, {Br{\"u}semeister},
  {Bucciarelli}, {Budnik}, {Burgess}, {Burgon}, {Burlacu}, {Busonero}, {Buzzi},
  {Caffau}, {Cambras}, {Campbell}, {Cancelliere}, {Cantat-Gaudin}, {Carlucci},
  {Carrasco}, {Castellani}, {Charlot}, {Charnas}, {Charvet}, {Chassat},
  {Chiavassa}, {Clotet}, {Cocozza}, {Collins}, {Collins}, {Costigan}, {Crifo},
  {Cross}, {Crosta}, {Crowley}, {Dafonte}, {Damerdji}, {Dapergolas}, {David},
  {David}, {De Cat}, {de Felice}, {de Laverny}, {De Luise}, {De March}, {de
  Martino}, {de Souza}, {Debosscher}, {del Pozo}, {Delbo}, {Delgado},
  {Delgado}, {di Marco}, {Di Matteo}, {Diakite}, {Distefano}, {Dolding}, {Dos
  Anjos}, {Drazinos}, {Dur{\'a}n}, {Dzigan}, {Ecale}, {Edvardsson}, {Enke},
  {Erdmann}, {Escolar}, {Espina}, {Evans}, {Eynard Bontemps}, {Fabre},
  {Fabrizio}, {Faigler}, {Falc{\~a}o}, {Farr{\`a}s Casas}, {Faye}, {Federici},
  {Fedorets}, {Fern{\'a}ndez-Hern{\'a}ndez}, {Fernique}, {Fienga}, {Figueras},
  {Filippi}, {Findeisen}, {Fonti}, {Fouesneau}, {Fraile}, {Fraser}, {Fuchs},
  {Furnell}, {Gai}, {Galleti}, {Galluccio}, {Garabato}, {Garc{\'\i}a-Sedano},
  {Gar{\'e}}, {Garofalo}, {Garralda}, {Gavras}, {Gerssen}, {Geyer}, {Gilmore},
  {Girona}, {Giuffrida}, {Gomes}, {Gonz{\'a}lez-Marcos},
  {Gonz{\'a}lez-N{\'u}{\~n}ez}, {Gonz{\'a}lez-Vidal}, {Granvik}, {Guerrier},
  {Guillout}, {Guiraud}, {G{\'u}rpide}, {Guti{\'e}rrez-S{\'a}nchez}, {Guy},
  {Haigron}, {Hatzidimitriou}, {Haywood}, {Heiter}, {Helmi}, {Hobbs},
  {Hofmann}, {Holl}, {Holland}, {Hunt}, {Hypki}, {Icardi}, {Irwin}, {Jevardat
  de Fombelle}, {Jofr{\'e}}, {Jonker}, {Jorissen}, {Julbe}, {Karampelas},
  {Kochoska}, {Kohley}, {Kolenberg}, {Kontizas}, {Koposov}, {Kordopatis},
  {Koubsky}, {Kowalczyk}, {Krone-Martins}, {Kudryashova}, {Kull}, {Bachchan},
  {Lacoste-Seris}, {Lanza}, {Lavigne}, {Le Poncin-Lafitte}, {Lebreton},
  {Lebzelter}, {Leccia}, {Leclerc}, {Lecoeur-Taibi}, {Lemaitre}, {Lenhardt},
  {Leroux}, {Liao}, {Licata}, {Lindstr{\o}m}, {Lister}, {Livanou}, {Lobel},
  {L{\"o}ffler}, {L{\'o}pez}, {Lopez-Lozano}, {Lorenz}, {Loureiro},
  {MacDonald}, {Magalh{\~a}es Fernandes}, {Managau}, {Mann}, {Mantelet},
  {Marchal}, {Marchant}, {Marconi}, {Marie}, {Marinoni}, {Marrese},
  {Marschalk{\'o}}, {Marshall}, {Mart{\'\i}n-Fleitas}, {Martino}, {Mary},
  {Matijevi{\v{c}}}, {Mazeh}, {McMillan}, {Messina}, {Mestre}, {Michalik},
  {Millar}, {Miranda}, {Molina}, {Molinaro}, {Molinaro}, {Moln{\'a}r},
  {Moniez}, {Montegriffo}, {Monteiro}, {Mor}, {Mora}, {Morbidelli}, {Morel},
  {Morgenthaler}, {Morley}, {Morris}, {Mulone}, {Muraveva}, {Musella},
  {Narbonne}, {Nelemans}, {Nicastro}, {Noval}, {Ord{\'e}novic},
  {Ordieres-Mer{\'e}}, {Osborne}, {Pagani}, {Pagano}, {Pailler}, {Palacin},
  {Palaversa}, {Parsons}, {Paulsen}, {Pecoraro}, {Pedrosa}, {Pentik{\"a}inen},
  {Pereira}, {Pichon}, {Piersimoni}, {Pineau}, {Plachy}, {Plum}, {Poujoulet},
  {Pr{\v{s}}a}, {Pulone}, {Ragaini}, {Rago}, {Rambaux}, {Ramos-Lerate},
  {Ranalli}, {Rauw}, {Read}, {Regibo}, {Renk}, {Reyl{\'e}}, {Ribeiro},
  {Rimoldini}, {Ripepi}, {Riva}, {Rixon}, {Roelens}, {Romero-G{\'o}mez},
  {Rowell}, {Royer}, {Rudolph}, {Ruiz-Dern}, {Sadowski}, {Sagrist{\`a}
  Sell{\'e}s}, {Sahlmann}, {Salgado}, {Salguero}, {Sarasso}, {Savietto},
  {Schnorhk}, {Schultheis}, {Sciacca}, {Segol}, {Segovia}, {Segransan},
  {Serpell}, {Shih}, {Smareglia}, {Smart}, {Smith}, {Solano}, {Solitro},
  {Sordo}, {Soria Nieto}, {Souchay}, {Spagna}, {Spoto}, {Stampa}, {Steele},
  {Steidelm{\"u}ller}, {Stephenson}, {Stoev}, {Suess}, {S{\"u}veges}, {Surdej},
  {Szabados}, {Szegedi-Elek}, {Tapiador}, {Taris}, {Tauran}, {Taylor},
  {Teixeira}, {Terrett}, {Tingley}, {Trager}, {Turon}, {Ulla}, {Utrilla},
  {Valentini}, {van Elteren}, {Van Hemelryck}, {van Leeuwen}, {Varadi},
  {Vecchiato}, {Veljanoski}, {Via}, {Vicente}, {Vogt}, {Voss}, {Votruba},
  {Voutsinas}, {Walmsley}, {Weiler}, {Weingrill}, {Werner}, {Wevers},
  {Whitehead}, {Wyrzykowski}, {Yoldas}, {{\v{Z}}erjal}, {Zucker}, {Zurbach},
  {Zwitter}, {Alecu}, {Allen}, {Allende Prieto}, {Amorim},
  {Anglada-Escud{\'e}}, {Arsenijevic}, {Azaz}, {Balm}, {Beck}, {Bernstein},
  {Bigot}, {Bijaoui}, {Blasco}, {Bonfigli}, {Bono}, {Boudreault}, {Bressan},
  {Brown}, {Brunet}, {Bunclark}, {Buonanno}, {Butkevich}, {Carret}, {Carrion},
  {Chemin}, {Ch{\'e}reau}, {Corcione}, {Darmigny}, {de Boer}, {de Teodoro}, {de
  Zeeuw}, {Delle Luche}, {Domingues}, {Dubath}, {Fodor}, {Fr{\'e}zouls},
  {Fries}, {Fustes}, {Fyfe}, {Gallardo}, {Gallegos}, {Gardiol}, {Gebran},
  {Gomboc}, {G{\'o}mez}, {Grux}, {Gueguen}, {Heyrovsky}, {Hoar}, {Iannicola},
  {Isasi Parache}, {Janotto}, {Joliet}, {Jonckheere}, {Keil}, {Kim},
  {Klagyivik}, {Klar}, {Knude}, {Kochukhov}, {Kolka}, {Kos}, {Kutka}, {Lainey},
  {LeBouquin}, {Liu}, {Loreggia}, {Makarov}, {Marseille}, {Martayan},
  {Martinez-Rubi}, {Massart}, {Meynadier}, {Mignot}, {Munari}, {Nguyen},
  {Nordlander}, {Ocvirk}, {O'Flaherty}, {Olias Sanz}, {Ortiz}, {Osorio},
  {Oszkiewicz}, {Ouzounis}, {Palmer}, {Park}, {Pasquato}, {Peltzer}, {Peralta},
  {P{\'e}turaud}, {Pieniluoma}, {Pigozzi}, {Poels}, {Prat}, {Prod'homme},
  {Raison}, {Rebordao}, {Risquez}, {Rocca-Volmerange}, {Rosen}, {Ruiz-Fuertes},
  {Russo}, {Sembay}, {Serraller Vizcaino}, {Short}, {Siebert}, {Silva},
  {Sinachopoulos}, {Slezak}, {Soffel}, {Sosnowska}, {Strai{\v{z}}ys}, {ter
  Linden}, {Terrell}, {Theil}, {Tiede}, {Troisi}, {Tsalmantza}, {Tur},
  {Vaccari}, {Vachier}, {Valles}, {Van Hamme}, {Veltz}, {Virtanen}, {Wallut},
  {Wichmann}, {Wilkinson}, {Ziaeepour}, \& {Zschocke}}]{2016A&A...595A...1G}
{Gaia Collaboration}, {Prusti}, T., {de Bruijne}, J.~H.~J., {et~al.} 2016,
  \aap, 595, A1

\bibitem[{{Hunter}(2007)}]{2007CSE.....9...90H}
{Hunter}, J.~D. 2007, Computing in Science and Engineering, 9, 90

\bibitem[{{Jacobs} {et~al.}(2002){Jacobs}, {Jones}, {Lanyi}, {Lowe}, {Naudet},
  {Resch}, {Steppe}, {Zhang}, {Ulvestad}, {Taylor}, {Sovers}, {Ma}, {Gordon},
  {Fey}, {Boboltz}, \& {Charlot}}]{2002ivsg.conf..350J}
{Jacobs}, C.~S., {Jones}, D.~L., {Lanyi}, G.~E., {et~al.} 2002, in
  International VLBI Service for Geodesy and Astrometry: General Meeting
  Proceedings, ed. N.~R. {Vandenberg} \& K.~D. {Baver}, 350

\bibitem[{{Kovalev} {et~al.}(2008){Kovalev}, {Lobanov}, {Pushkarev}, \&
  {Zensus}}]{2008A&A...483..759K}
{Kovalev}, Y.~Y., {Lobanov}, A.~P., {Pushkarev}, A.~B., \& {Zensus}, J.~A.
  2008, \aap, 483, 759

\bibitem[{{Kovalev} {et~al.}(2017){Kovalev}, {Petrov}, \&
  {Plavin}}]{2017A&A...598L...1K}
{Kovalev}, Y.~Y., {Petrov}, L., \& {Plavin}, A.~V. 2017, \aap, 598, L1

\bibitem[{{Kovalev} {et~al.}(2020){Kovalev}, {Zobnina}, {Plavin}, \&
  {Blinov}}]{2020MNRAS.493L..54K}
{Kovalev}, Y.~Y., {Zobnina}, D.~I., {Plavin}, A.~V., \& {Blinov}, D. 2020,
  \mnras, 493, L54

\bibitem[{{Liu} {et~al.}(2020){Liu}, {Lambert}, {Zhu}, \&
  {Liu}}]{2020A&A...634A..28L}
{Liu}, N., {Lambert}, S.~B., {Zhu}, Z., \& {Liu}, J.~C. 2020, \aap, 634, A28

\bibitem[{{Makarov} {et~al.}(2012){Makarov}, {Berghea}, {Boboltz}, {Dieck},
  {Dorland}, {Dudik}, {Fey}, {Gaume}, {Lei}, {Schmitt}, \&
  {Zacharias}}]{2012MmSAI..83..952M}
{Makarov}, V., {Berghea}, C., {Boboltz}, D., {et~al.} 2012, \memsai, 83, 952

\bibitem[{{Makarov} {et~al.}(2019){Makarov}, {Berghea}, {Frouard}, {Fey}, \&
  {Schmitt}}]{2019ApJ...873..132M}
{Makarov}, V.~V., {Berghea}, C.~T., {Frouard}, J., {Fey}, A., \& {Schmitt},
  H.~R. 2019, \apj, 873, 132

\bibitem[{{Makarov} {et~al.}(2017){Makarov}, {Frouard}, {Berghea}, {Rest},
  {Chambers}, {Kaiser}, {Kudritzki}, \& {Magnier}}]{2017ApJ...835L..30M}
{Makarov}, V.~V., {Frouard}, J., {Berghea}, C.~T., {et~al.} 2017, \apjl, 835,
  L30

\bibitem[{{Malkin}(2018)}]{2018ApJS..239...20M}
{Malkin}, Z. 2018, \apjs, 239, 20

\bibitem[{{Mignard} \& {Klioner}(2012)}]{2012A&A...547A..59M}
{Mignard}, F. \& {Klioner}, S. 2012, \aap, 547, A59

\bibitem[{{Mignard} {et~al.}(2016){Mignard}, {Klioner}, {Lindegren}, {Bastian},
  {Bombrun}, {Hern{\'a}ndez}, {Hobbs}, {Lammers}, {Michalik}, {Ramos-Lerate},
  {Biermann}, {Butkevich}, {Comoretto}, {Joliet}, {Holl}, {Hutton}, {Parsons},
  {Steidelm{\"u}ller}, {Andrei}, {Bourda}, \& {Charlot}}]{2016A&A...595A...5M}
{Mignard}, F., {Klioner}, S., {Lindegren}, L., {et~al.} 2016, \aap, 595, A5

\bibitem[{Nadaraya(1964)}]{nadaraya1964estimating}
Nadaraya, E.~A. 1964, Theory of Probability \& Its Applications, 9, 141

\bibitem[{{Ochsenbein} {et~al.}(2000){Ochsenbein}, {Bauer}, \&
  {Marcout}}]{2000A&AS..143...23O}
{Ochsenbein}, F., {Bauer}, P., \& {Marcout}, J. 2000, \aaps, 143, 23

\bibitem[{{Orosz} \& {Frey}(2013)}]{2013A&A...553A..13O}
{Orosz}, G. \& {Frey}, S. 2013, \aap, 553, A13

\bibitem[{{Petrov} \& {Kovalev}(2017{\natexlab{a}})}]{2017MNRAS.471.3775P}
{Petrov}, L. \& {Kovalev}, Y.~Y. 2017{\natexlab{a}}, \mnras, 471, 3775

\bibitem[{{Petrov} \& {Kovalev}(2017{\natexlab{b}})}]{2017MNRAS.467L..71P}
{Petrov}, L. \& {Kovalev}, Y.~Y. 2017{\natexlab{b}}, \mnras, 467, L71

\bibitem[{{Petrov} {et~al.}(2019){Petrov}, {Kovalev}, \&
  {Plavin}}]{2019MNRAS.482.3023P}
{Petrov}, L., {Kovalev}, Y.~Y., \& {Plavin}, A.~V. 2019, \mnras, 482, 3023

\bibitem[{{Plavin} {et~al.}(2019){Plavin}, {Kovalev}, \&
  {Petrov}}]{2019ApJ...871..143P}
{Plavin}, A.~V., {Kovalev}, Y.~Y., \& {Petrov}, L.~Y. 2019, \apj, 871, 143

\bibitem[{{Porcas}(2009)}]{2009A&A...505L...1P}
{Porcas}, R.~W. 2009, \aap, 505, L1

\bibitem[{{Shabala} {et~al.}(2014){Shabala}, {Rogers}, {McCallum}, {Titov},
  {Blanchard}, {Lovell}, \& {Watson}}]{2014JGeod..88..575S}
{Shabala}, S.~S., {Rogers}, J.~G., {McCallum}, J.~N., {et~al.} 2014, Journal of
  Geodesy, 88, 575

\bibitem[{{Souchay} {et~al.}(2019){Souchay}, {Gattano}, {Andrei}, {Souami},
  {Coelho}, {Barache}, {Taris}, {Secrest}, \&
  {Berthereau}}]{2019A&A...624A.145S}
{Souchay}, J., {Gattano}, C., {Andrei}, A.~H., {et~al.} 2019, \aap, 624, A145

\bibitem[{{Taylor}(2011)}]{2011ascl.soft01010T}
{Taylor}, M. 2011, {TOPCAT: Tool for OPerations on Catalogues And Tables}

\bibitem[{Watson(1964)}]{watson1964smooth}
Watson, G.~S. 1964, Sankhy{\=a}: The Indian Journal of Statistics, Series A,
  359

\bibitem[{{Xu} {et~al.}(2019){Xu}, {Anderson}, {Heinkelmann}, {Lunz}, {Schuh},
  \& {Wang}}]{2019ApJS..242....5X}
{Xu}, M.~H., {Anderson}, J.~M., {Heinkelmann}, R., {et~al.} 2019, \apjs, 242, 5

\bibitem[{{Xu} {et~al.}(2021){Xu}, {Lunz}, {Anderson}, {Savolainen}, {Zubko},
  \& {Schuh}}]{2021A&A...647A.189X}
{Xu}, M.~H., {Lunz}, S., {Anderson}, J.~M., {et~al.} 2021, \aap, 647, A189

\bibitem[{{Zacharias} \& {Zacharias}(2014)}]{2014AJ....147...95Z}
{Zacharias}, N. \& {Zacharias}, M.~I. 2014, \aj, 147, 95

\end{thebibliography}



\end{document}